\documentclass[11pt,twoside,letterpaper]{article} 
\usepackage{times,fancyhdr}
\usepackage[dvips]{graphicx}

\usepackage{epsfig}
\usepackage{color}
\usepackage{graphicx}
\usepackage{amsmath}
\usepackage{amsfonts}
\usepackage{amssymb}


\makeatletter
\def\cleardoublepage{\clearpage\if@twoside \ifodd\c@page\else%
    \hbox{}%
    \thispagestyle{empty}%
    \newpage%
    \if@twocolumn\hbox{}\newpage\fi\fi\fi}
\makeatother

\def\figurename{Figure}
\makeatletter
\renewcommand{\fnum@figure}[1]{\figurename~\thefigure.}
\makeatother

\def\tablename{Table}
\makeatletter
\renewcommand{\fnum@table}[1]{\tablename~\thetable.}
\makeatother

\begin{document}

\title{
{\begin{flushleft}
\vskip 0.45in
\end{flushleft}
\vskip 0.45in \bfseries\scshape Environmental Noise and Nonlinear Relaxation in Biological Systems}}
\author{\bfseries\itshape B. Spagnolo$^1$\thanks{E-mail address: bernardo.spagnolo@unipa.it}, 
D. Valenti$^1$\thanks{E-mail address: davide.valenti@unipa.it}, 
S. Spezia$^1$\thanks{E-mail address: stefano.spezia@gmail.com}, 
L. Curcio$^2$, \\
\bfseries\itshape N. Pizzolato$^1$\thanks{E-mail address: nicola.pizzolato@unipa.it}, 
A. A. Dubkov$^3$\thanks{E-mail address: dubkov@rf.unn.ru}, 
A. Fiasconaro$^{1,4}$, 
D. Persano Adorno$^1$, \\
\bfseries\itshape P. Lo Bue$^5$, 
E. Peri$^5$ and  
S. Colazza$^5$\\
$^1$\,Dipartimento di Fisica, \\
    Group of Interdisciplinary Physics\\
    Universit\`{a} di Palermo and CNISM-INFM, Unit\`{a} di Palermo\\
    Viale delle Scienze, ed. 18, I-90128 Palermo, Italy\\
$^2$\,Dipartimento di Ingegneria Elettrica, Elettronica e delle
    Telecomunicazioni, \\
    Viale delle Scienze, ed. 9, I-90128 Palermo,
    Italy\\
$^3$\,Radiophysics Department, Nizhniy Novgorod State University, \\
    23 Gagarin ave.,  603950 Nizhniy Novgorod, Russia\\
$^4$\,Departamento de F\'{\i}sica de la Materia Condensada,
    Universidad de Zaragoza, \\
    E-50009 Zaragoza, Spain\\
$^5$\,Dipartimento di Scienze Entomologiche, Fitopatologiche,\\
    Microbiologiche, Agrarie e Zootecniche, Universit\`{a} di Palermo,\\
    Viale delle Scienze, ed. 5, I-90128 Palermo, Italy
}

\date{}
\maketitle
\thispagestyle{empty}
\setcounter{page}{1}
\thispagestyle{fancy}
\fancyhead{}
\fancyfoot{}
\renewcommand{\headrulewidth}{0pt}

\begin{abstract}
We analyse the effects of environmental noise in three different
biological systems: (i) mating behaviour of individuals of
\emph{Nezara viridula} (L.) (Heteroptera Pentatomidae); (ii) polymer translocation in crowded
solution; (iii) an ecosystem described by a Verhulst model with a
multiplicative L\'{e}vy noise. Specifically, we report on
experiments on the behavioural response of \emph{N. viridula}
individuals to sub-threshold deterministic signals in the presence
of noise. We analyse the insect response by directionality tests
performed on a group of male individuals at different noise
intensities. The percentage of insects which react to the
sub-threshold signal shows a non-monotonic behavior, characterized
by the presence of a maximum, for increasing values of the noise
intensity. This is the signature of the non-dynamical stochastic
resonance phenomenon. By using a "hard" threshold model we find that
the maximum of the signal-to-noise ratio occurs in the same range
of noise intensity values for which the behavioral activation shows
a maximum. In the second system, the noise driven translocation of
short polymers in crowded solutions is analyzed. An improved version
of the Rouse model for a flexible polymer has been adopted to mimic
the molecular dynamics, by taking into account both the interactions
between adjacent monomers and introducing a Lennard-Jones potential
between non-adjacent beads. A bending recoil torque has also been
included in our model. The polymer dynamics is simulated in a
two-dimensional domain by numerically solving the Langevin equations
of motion. Thermal fluctuations are taken into account by
introducing a Gaussian uncorrelated noise. The mean first
translocation time of the polymer centre of inertia shows a minimum
as a function of the frequency of the oscillating forcing field. In
the third ecosystem, the transient dynamics of the Verhulst model
perturbed by arbitrary non-Gaussian white noise is investigated.
Based on the infinitely divisible distribution of the L\'{e}vy
process we study the nonlinear relaxation of the population density
for three cases of white non-Gaussian noise: (i) shot noise, (ii)
noise with a probability density of increments expressed in terms of
Gamma function, and (iii) Cauchy stable noise. We obtain exact
results for the probability distribution of the population density
in all cases, and for Cauchy stable noise the exact expression of
the nonlinear relaxation time is derived. Moreover starting from an
initial delta function distribution, we find a transition induced by
the multiplicative L\'{e}vy noise, from a trimodal probability
distribution to a bimodal probability distribution in asymptotics.
Finally we find a nonmonotonic behavior of the nonlinear relaxation
time as a function of the Cauchy stable noise intensity.

\end{abstract}


\pagestyle{fancy}
\fancyhead{}
\fancyhead[EC]{B. Spagnolo, D. Valenti, S. Spezia et al.}
\fancyhead[EL,OR]{\thepage}
\fancyhead[OC]{Environmental Noise and Nonlinear Relaxation in Biological System}
\fancyfoot{}
\renewcommand\headrulewidth{0.5pt}

\section{Introduction}
\label{intro}\vskip-0.2cm
During last decades noise-induced effects have been experimentally
observed and theoretically studied in different physical and
biological
contexts~\cite{Agu01,Spa03,Val04,Zim99,Bjo01,Gre98,Chi08,Giu09,Piz09a},
such as neuronal cells, excitable systems and threshold physical
systems~\cite{Bra94,Mos94,Gin95,Pei95,Pik97,Noz98,Lon98,Sto00,Wie94,Gam95,Wan00,Lin04}.

In particular, stochastic resonance, resonant activation and noise
enhanced stability phenomena in neuronal activation have been
recently discussed~\cite{Lin04,Dua08,Pol05}.

Nature consists of open systems characterized by intrinsically
non-linear interactions and subject to environmental
noise~\cite{Spa04}. The presence of random fluctuations, that are an
uneliminable component of natural ecosystems, makes difficult
detection and transmission of signals and can modify the information
transported.

However, in the presence of some specific non-linearity of the
system and for suitable intensity of noise, counterintuitive
phenomena, such as stochastic resonance (SR), can be observed. This
indicates that noise can play a constructive role, improving the
conditions for signal detection.

SR phenomenon initially was observed in the temperature cycles of
the Earth~\cite{Benzi}, can be found in many physical and biological
non-linear systems~\cite{Gam98,Man94,Agu10}. SR can be modelled by a
bistable potential subject to periodical driving force in the
presence of external additive noise. The signature of SR is a
non-monotonic behaviour, characterized by a maximum, of the
signal-to-noise (SNR) ratio as a function of the noise intensity.
This indicates that the noise can enhance the amplitude of
deterministic signals, improving the response of the system through
a resonance-like
phenomenon~\cite{Mos94,Gin95,Pei95,Noz98,Lon98,Sto00,Wie94,Gam95,Wan00,
Lin04,Vil98,Bul91,Nei02,Bah02,Dou93,Rus99,Fre02,Gre00,Gam98,Man94,Gai97}.
However, SR does not occur only in bistable systems, but also in
monostable, excitable, and non-dynamical systems. In these
situations we name this effect non-dynamical (or threshold)
stochastic resonance, because the phenomenon is connected with the
crossing of a threshold and can occur also in the absence of an
external potential~\cite{Mos94,Gin95,Vil98}. Sensory neurons, that
are threshold systems characterized by intrinsic noise, are an ideal
workbench to observe non-dynamical SR
because~\cite{Bul91,Nei02,Bah02}. Historical experiments revealed
the presence of non-dynamical SR in the neural response of
mechanoreceptor cells of crayfish~\cite{Dou93}, and the improvement
of sensorial activity of paddlefish in the detection of electric
signals produced by preys~\cite{Rus99,Fre02,Gre00}. Such sensory
neurons are ideally suited to exhibit SR as they are intrinsically
noisy and operate as threshold systems~\cite{Bul91,Nei02,Bah02}.

In this contribution, we study the effects of external noise in
three different biological systems. We start analyzing the mating
behaviour of individuals of \emph{N. viridula} (L.) (Heteroptera
Pentatomidae). In particular, we investigate the role of noise in
the response of male insects to mechanical vibrations emitted by
female individuals and transmitted in the
substrate~\cite{Cok99,Cok03,Cok07}. \emph{N. viridula}, the southern
green stink bug, is a pentatomid insect highly polyphagous and quite
harmful for agricolture~\cite{Tod89,Pan00}. \emph{N. viridula} has
up to five generations per year~\cite{Bor87,Kir64,Tre81,Fuc03}.

The mating behavior of \emph{N. viridula} can be divided into
long-range location and short-range courtship. The first one
includes those components of the behavior that lead to the arrival
of females in the vicinity of males. The long range attraction
mediated by male attractant pheromone enables both sexes to reach
the same plant.

Here, we analyze the mating behaviour of insects during the
short-range courtship, when bugs of both sexes are very close and
the acoustic stimuli (improperly called songs) can be an important
element in the sexual communication~\cite{Cok99}.

The sound is produced by the tymbal, an organ sited in the back and
present in adult individuals~\cite{Cok03}. The vibrations, produced
by a bug at the frequency of about $100$ Hz, propagate through the legs
into the plant stem and can be detected by the vibro-receptors
placed in the legs of another insect~\cite{Tre81,Bag08}. Many
experimental studies have been performed on this acoustic
communication, analysing the different signals characteristic of
populations of \emph{N. viridula} from Slovenia, Florida, Japan and
Australia~\cite{Cok00}.

The fundamental role of the vibratory signals suggests that a better
knowledge of the mechanism of acoustic communication during the
short-range courtship can help to point out more efficient strategy
to control \emph{N. viridula} populations, devising "biologic" traps
whose working principle is the emission of acoustic signals. In
natural conditions, \emph{N. viridula} populations interact strongly
with environment, and therefore the presence of surrounding noise
becomes an essential component of the acoustic communication.

In the second part of this contribution, we consider transport
phenomena of polymers in crowded solutions. In fact, the
translocation of DNA and RNA across nuclear membranes as well as the
crossing of potential barriers by many proteins represents a
fundamental process in cellular biology. The study of the transport
of macromolecules across nanometer size channels is important for
both medical research in anticancer targeted therapy
\cite{Higgins2007,Hal2009} and technological applications
\cite{Mann06,Sunda2008}.

First experiments on the passage of DNA molecules across an
$\alpha$-hemolysin ($\alpha$-HL) protein channel revealed a linear
relationship of the most probable crossing time $\tau_p$ with the
molecule length \cite{Kas96}. Moreover, $\tau_p$ scales as the
inverse square of the temperature and the dynamics of biopolymer
translocation across an $\alpha$-HL channel is found to be governed
by pore-molecule interactions \cite{Ake99,Mel00,Mel02}. More recent
experimental studies have shown that the application of an AC
voltage to drive the translocation process of DNA molecules through
a nanopore plays a significant role in the DNA-nanopore interaction,
and provides new insights into the DNA conformations
\cite{Den03,Ver04,Sig08,Lat09,Nik09}.

The complex scenario of the translocation dynamics coming from
experiments has been enriched by several theoretical and simulative
studies \cite{Lub99,Sto05,For07,Luo08,Gra08,Piz08,Pan08,Piz09}. The
mean first passage time of a Brownian particle to cross a potential
barrier in the presence of thermal fluctuations and a periodic
forcing field has been theoretically and experimentally investigated
as a function of the driving frequency in Refs.
\cite{Doe92,Bie93,Bog98,Man00,Dub04,Spa07}. The translocation time
of chain polymers has been theoretically studied in the presence of
a dichotomically fluctuating chemical potential only as a function
of its amplitude in Ref.~\cite{Park98}.

In particular we investigate the role of an external oscillating
forcing field on the transport dynamics of short polymers
surmounting a barrier, in the presence of a metastable state. We
find a minimum of the mean first translocation time (MFTT) of the
molecule center of mass as a function of the frequency of the
forcing field. This nonlinear behaviour represents the resonant
activation (RA) phenomenon in polymer translocation. We find that a
suitable tuned oscillating field can speed up or slow down the mean
time of the translocation process of a molecule crossing a barrier,
using the frequency as a control parameter. This effect can be of
fundamental importance for all those experiments on cell metabolism,
DNA-RNA sequencing and drug delivery mechanism in anti-cancer
therapy.

In the third part of this chapter we investigate the transient
dynamics of the Verhulst model perturbed by arbitrary non-Gaussian
white noise. The nonlinear stochastic systems with noise excitation
have attracted extensive attention and the concept of noise-induced
transitions has got a wide variety of applications in physics,
chemistry, and biology~\cite{Hor84}. Noise-induced transitions are
conventionally defined in terms of changes in the number of extrema
in the probability distribution of a system variable and may depend
both quantitatively and qualitatively on the character of the noise,
i.e. on the properties of stochastic process which describes the
noise excitation. The Verhulst model, which is a cornerstone of
empirical and theoretical ecology, is one of the classic examples of
self-organization in many natural and artificial
systems~\cite{Eig79}. This model, also known as the logistic model,
is relevant to a wide range of situations including population
dynamics~\cite{Hor84,Mor82,Ciu93,Mat00}, self-replication of
macromolecules~\cite{Eig71}, spread of viral epidemics~\cite{Ace06},
cancer cell population \cite{Bao03}, biological and biochemical
systems~\cite{Der90,Ciu96}, population of photons in a single mode
laser~\cite{McN74,Oga83}, autocatalytic chemical
reactions~\cite{Sch72,Cha76,Gar77,Bou82,Leu87}, freezing of
supercooled liquids~\cite{Das83}, social
sciences~\cite{Her72,Mon78}, etc.

By considering the season fluctuations and the random availability
of resources we analyze the stochastic Verhulst equation in the
presence of a non-Gaussian stochastic process.  By investigating the
transient dynamics of this model we obtain exact results for the
mean value of the population density and its nonstationary
probability distribution for different types of white non-Gaussian.
Noise-induced transitions for the probability distribution of the
population density and a nonmonotonic behavior of the nonlinear
relaxation time as a function of the Cauchy noise intensity are
found.

The chapter is organized as follows. In section~\ref{sec:2} we
report on experimental setup and methods used in the investigation
of behavioural response in \emph{N. viridula}. In
section~\ref{sec:3}, we present our experimental results of
directionality tests on the behaviour of male individuals of
\emph{N. viridula}. In section~\ref{sec:4} we discuss the
experimental findings and compare them with theoretical results
obtained by a hard threshold model.

In Sect.~\ref{mod} we present our polymer chain model and give the
details of the molecular dynamics simulation process. Results are
reported in Sect.~\ref{res}. In the next section $IV$ we present our
Verhulst stochastic model with L\'{e}vy noise excitation together
with all the theoretical results obtained. Finally conclusions are
drawn in Sect.~\ref{conc}.

\section{Behavioural Response in \emph{N. viridula}}
\subsection{Materials and methods}
\label{sec:2} In our experiments we used individuals of \emph{N.
viridula} collected in the countryside around Palermo, and reared in
laboratory conditions~\cite{Col04}. Male insects have been used for
experimental trials after they reached sexual maturity (not less
than ten days after the final moult), and a three-day period of
isolation from the opposite sex~\cite{Cok07,Cok00}.

The sexual calling song emitted from a female individual has been
recorded by the membrane of a conic low-middle frequency loudspeaker
(MONACOR SPH 165 C CARBON with a diameter of $16.5 \thinspace cm$).
Afterwards the sound, stored on a pc, has been analysed and
processed using a commercial software. The speaker has been used as
an "inverse" microphone, namely an acoustic-electric transducer: the
sounds have been recorded from a low-frequency non-resonating
membrane of a speaker, conveniently chosen to get a good frequency
response starting at $20 \thinspace$ Hz. The sound acquisitions have
been made inside an anechoic chamber (sound insulated) at
$22-26^\circ$C, $70-80\%$ of relative moisture and in presence of
artificial light. The choice of this recording set-up has been
decided after a comparative analysis with a recording system based
on the use of a stethoscope. In particular, the speaker membrane
shows greater sensitivity at medium-low frequencies, that are
crucial to our experiment.

The sound has been sampled from the analogical signal source
($44100$ samples per second at $16$-bit) and then filtered by an
$18^{th}$ order Tchebychev filter (type I) with band-pass from $60$
to $400$ Hz. This filtering has been done to cut: (i) the low
frequencies due to the electric network ($50 Hz$) and those from the
conic loudspeaker, and (ii) the high frequencies due to the
electronic apparatus. Spectral and temporal properties of the
measured non-pulsed female calling songs (NPFCS) have been compared
with those of North America, observing that \emph{N. viridula}
individuals collected in Sicily have the same dialect as adults of
\emph{N. viridula} collected in USA with a slightly different
frequency range \cite{Cok07,Cok00,Cok05}.

\begin{figure*}
\begin{center}
\resizebox{1.00\columnwidth}{!}{%
\includegraphics{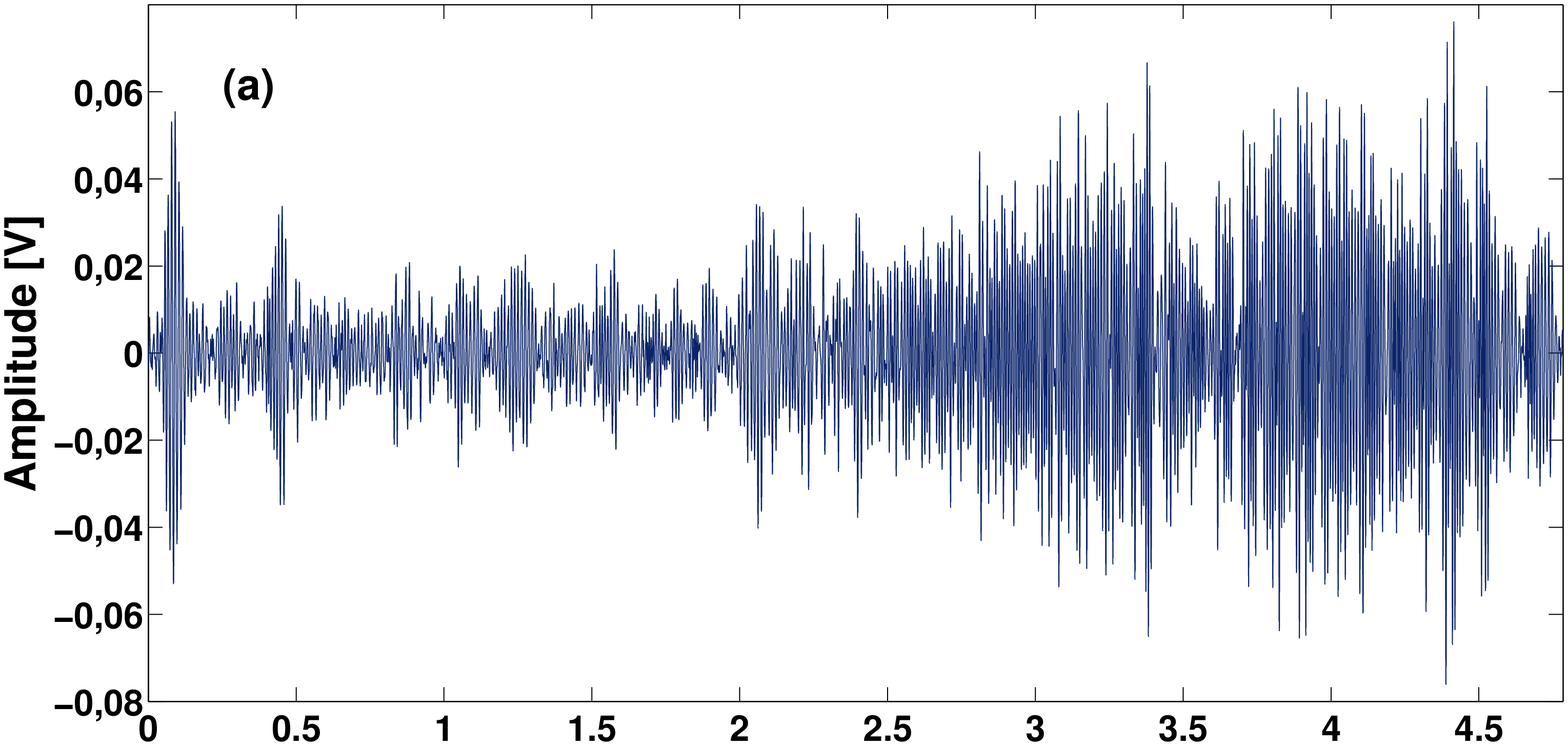}
\includegraphics{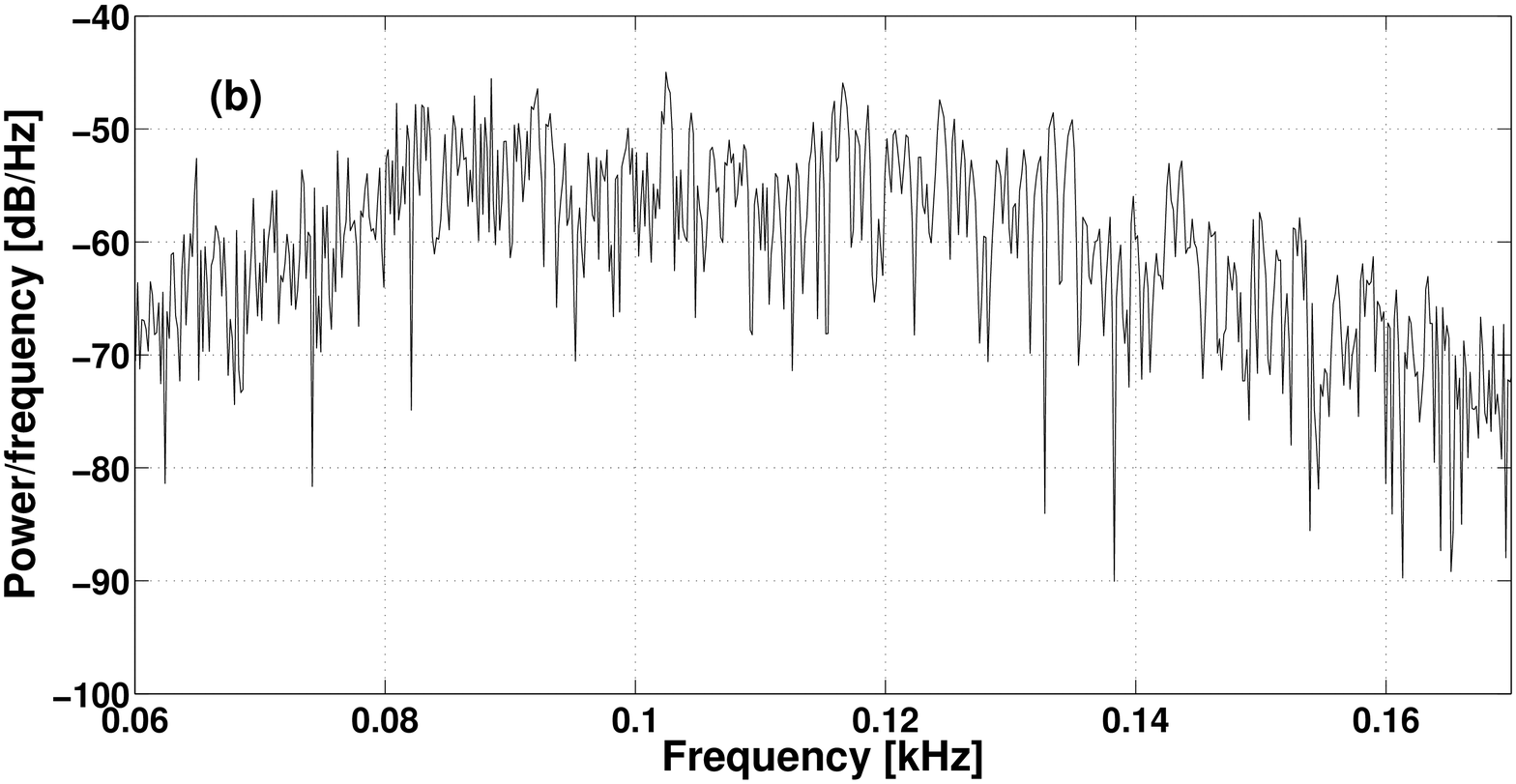}
}
\resizebox{1.00\columnwidth}{!}{%
\includegraphics{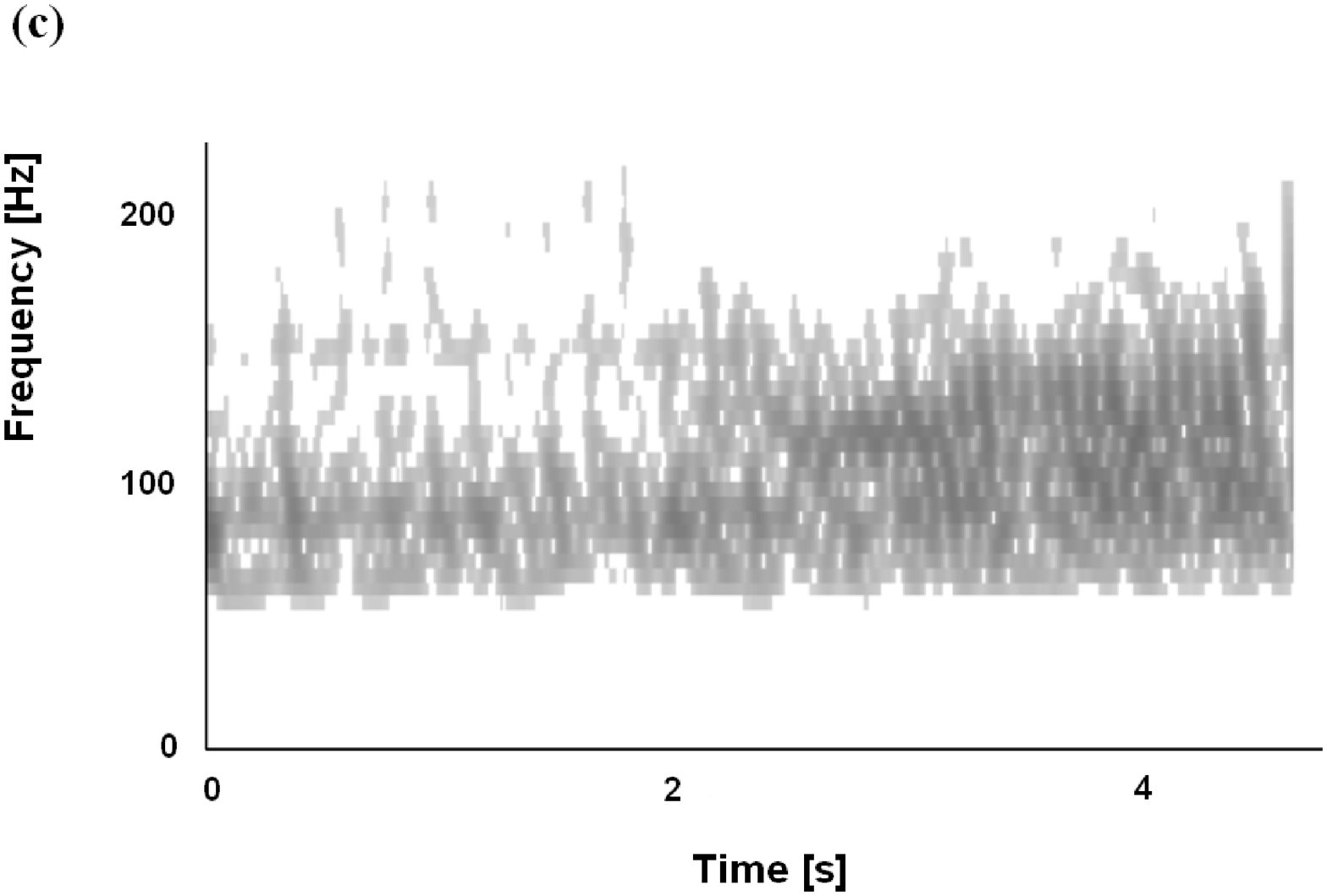}
\includegraphics{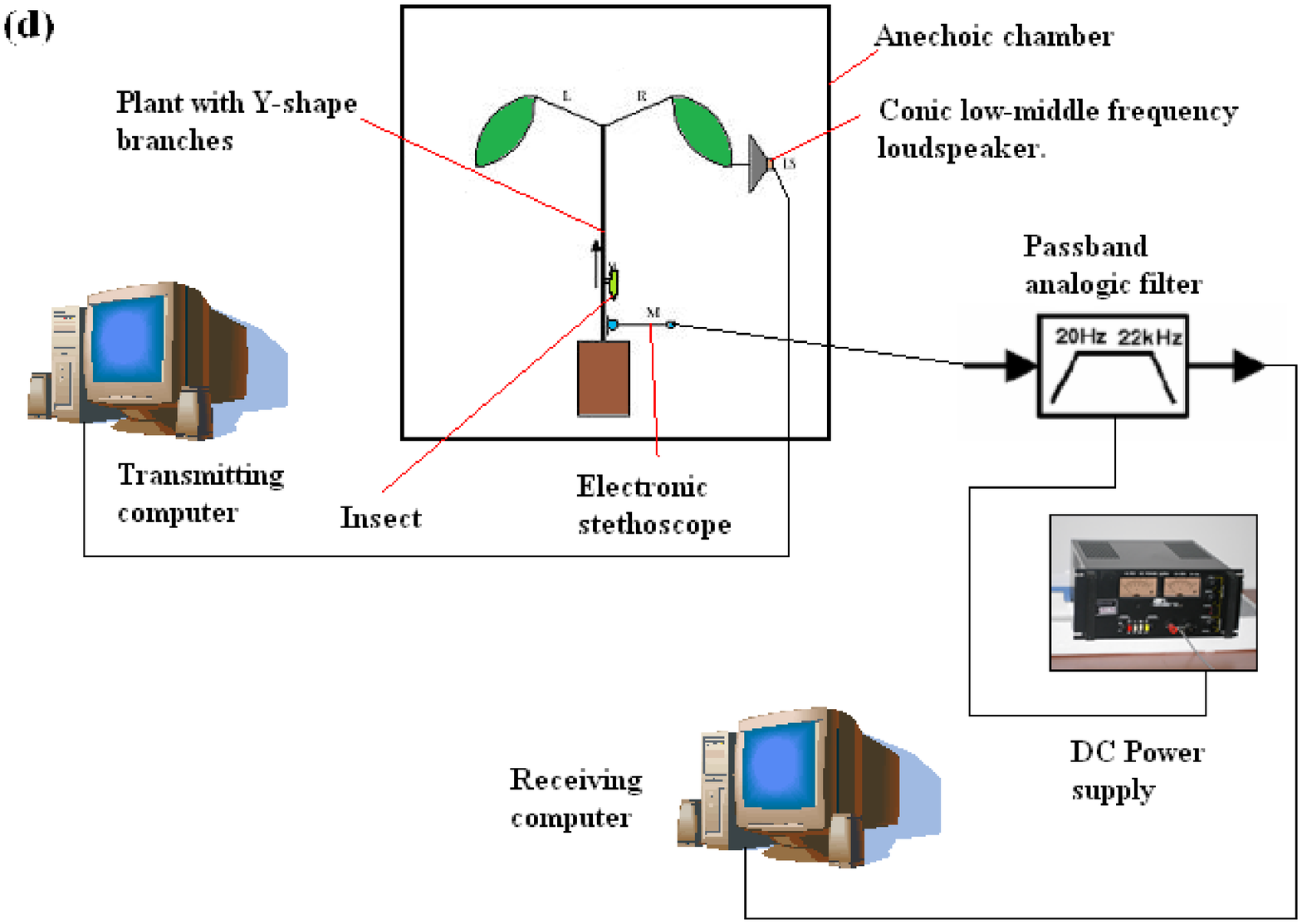}
}
\caption{(a) Oscillogram (b) Power spectrum density and (c) Sonagram of the
non pulsed type of \emph{Nezara viridula} female calling song; (d)
The block diagram of the experimental setup.}
\label{fig:1}       
\end{center}
\end{figure*}

In Fig.~\ref{fig:1}a, the oscillogram of NPFCS is shown. The signal
is characterized by a short pre-pulse followed by a longer one,
according to previous experimental findings~\cite{Cok00}. In
Fig.~\ref{fig:1}b, the power spectrum density (PSD) of NPFCS is
shown. In this spectrum the dominant frequencies range from $70$ to
$170$ Hz and the subdominant peaks do not exceed $400$ Hz. The
maximum peak occurs at $102.5$ Hz. In Fig.~\ref{fig:1}c we report
the relative sonagram, achieved by the Short Time Fourier Transform
(STFT) method. The STFT maps a signal providing information both
about frequencies and occurrence times. It shows that during the
first two seconds (short pre-pulse) the dominant frequency interval
is narrower than the range observed in the subsequent time space. In
particular in the first time interval the highest frequency does not
exceed $130$ Hz, whereas in the final one it reaches almost $170$~Hz.


Here we study the effects of noise on the behavior of \emph{N.
viridula} during the mating period. Therefore, in order to perform
directionality tests, we have designed and constructed a Y-shaped
dummy plant, and placed it inside an anechoic chamber. The Y-shaped
plant consists of a vertical wood stem, 10 cm long, and 0.8 - 0.9 cm
thick at the top of which there are two identical wooden branches,
25 cm long, and 0.4 cm thick, as shown in Fig.~\ref{fig:1}d. The
angle between two branches is $30^\circ-50^\circ$.

In our experiment a signal is sent along one branch of the Y-shaped
substrate and the behavior of single male individuals, initially
placed at the center of the vertical stem, is
observed~\cite{Cok99} (see Fig.~\ref{fig:1}d). The source of
vibratory signals (i.e. the cone used as an electro-acoustic
transducer) is in contact with the right apex of the Y-shaped dummy
plant. Vertical stem and lateral branches are not in direct contact,
albeit in close (0.5 cm) proximity.

We consider a trial valid if the insect, before choosing one
direction in the Y-shaped structure, has checked the two possible
directions of signal origin, touching the lower extremity of both
branches. By following these criteria, we made several observations
for different intensities of female calling songs, recording the
choices (left or right) of each male individual used in our trials,
and obtaining a set of statistical data that allows us to determine
the intensity threshold value at which the bugs start to "hear" the
calling song.

\subsection{Experimental results} \label{sec:3} 
\begin{figure*}[hbt]
\begin{center}
\resizebox{1.0\columnwidth}{!}{%
\includegraphics{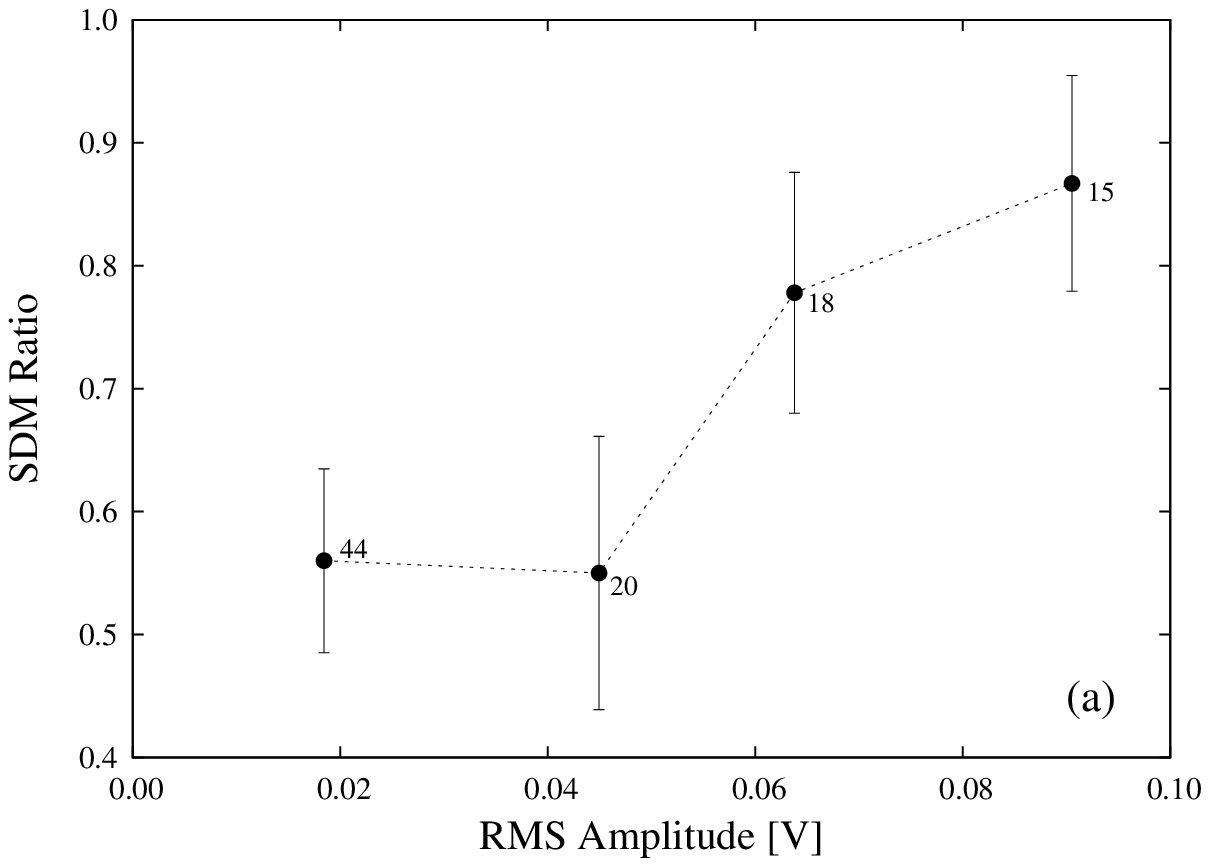}
\includegraphics{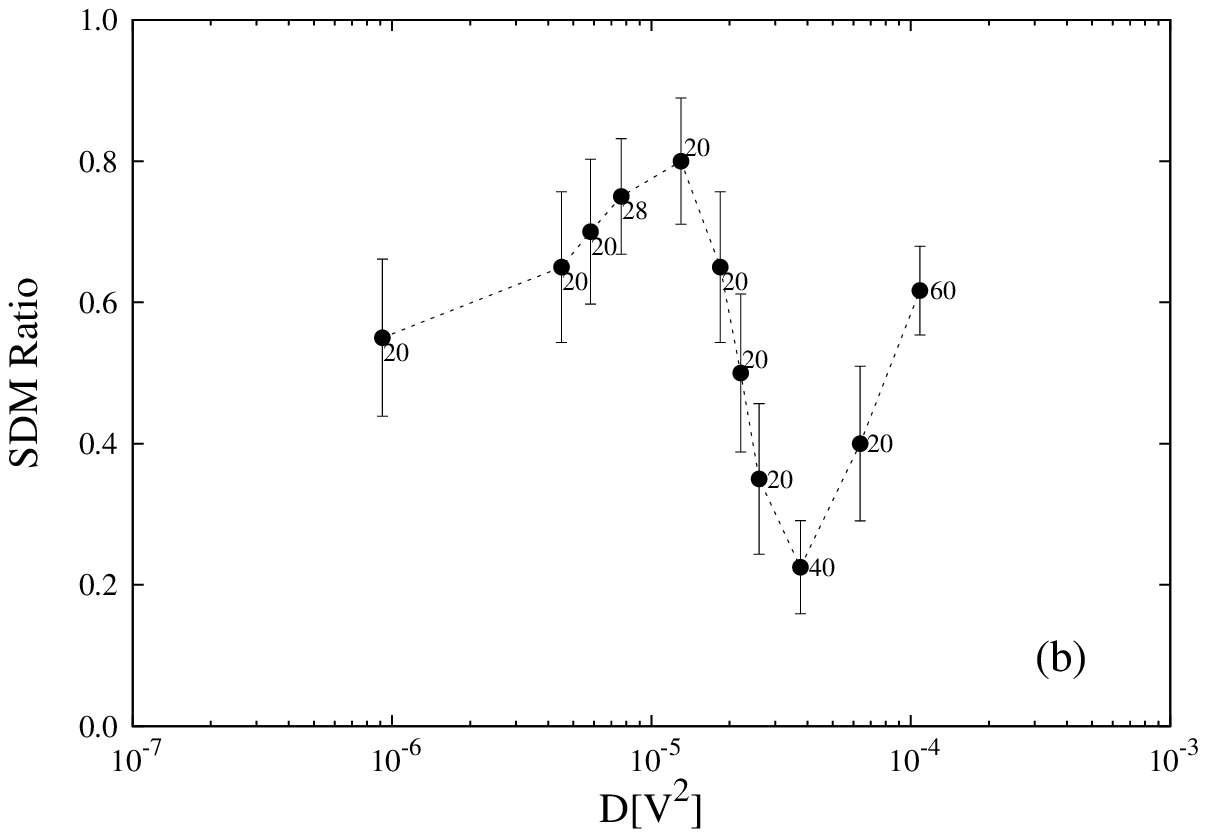}
}
\caption{Plots of the \emph{Source-Direction Movement (SDM) Ratio}
as a function of: (a) the female calling song Root Mean Square (RMS)
amplitude (purely deterministic signal); (b) the noise intensity
\emph{D}. In each experimental value is reported the error bar and
beside the corresponding number of the performed trials.}
\label{fig:2}       
\end{center}
\end{figure*}

The presence of an
"oriented" behavior, that is the tendency of the insects to choose the branch
with the signal source, is revealed by performing directionality
tests on a group of male individuals. When we observe a percentage
of insects higher than $65\%$ going towards the acoustic source,
\emph{source-direction movement} (SDM), we consider that the signal
has been revealed by the insects. In Fig.~\ref{fig:2}a we plot the
relative frequency of SDMs, that is the number of SDMs divided by
the total trials, at different signal intensities. The exact number
of trials, performed for each intensity, is reported beside the
corresponding point in the graph. For small values (lower than
0.0010 V) of the signal power approximately $50\%$ of the insects
choose one direction and the remaining $50\%$ the other.
%
Conversely, for values greater than 0.0020 V, the insects show a
preferential behavior, choosing the direction from which the signal
originates in the $80\%$ of the trials. Consequently, we have chosen
the value 0.0015 V of the signal power as the \emph{threshold level}
for signal detection.

Then, by using a sub-threshold signal plus a Gaussian white noise
signal we have investigated the response of the test insect for
different levels of noise intensity $D$. In Fig.~\ref{fig:2}b we
report the percentage of SDMs as a function of $D$, finding the
optimal noise intensity that maximizes the recognition between
individuals of opposite sex. The graph shows a maximum for $D\approx
1.30 \cdot 10^{-5}$ $V^2$. For values of $D$ both lower and higher
than $1.30 \cdot 10^{-5}$ $V^2$, the response of insects is not
significant. In particular, for $D>1.30 \cdot 10^{-5}$ $V^2$ the
percentage of individuals going towards the acoustic source
decreases below $0.5$ reaching $0.2$ for $D\approx 3.75 \cdot
10^{-5}$ $V^2$. The other values of the SDM ratio close to $50\%$,
indicate that individuals of \emph{N. viridula} randomly choose the
direction of their motion, that is no oriented behavior occurs. The
non-monotonic behavior of SDM, with a maximum at $D\approx 1.30
\cdot 10^{-5}$ $V^2$, indicates that in the presence of a
sub-threshold deterministic signal, the environmental noise can play
a constructive role, amplifying the weak input signal and
contributing to improve the communication among individuals of
\emph{N. viridula}. The occurrence of a minimum in the SDM behavior
at $D\approx 3.75 \cdot 10^{-5}$ $V^2$, will be subject of further
investigations. A possible conjectural explanation is the following:
when the noise intensity is so large that the signal received from
the vibro-receptors is significantly modified, the male insects are
not able to recognize the female calling song, and they exchange it
for the song of some rivals.

A further increase of the noise intensity causes the spectrum of the
received signal to become indistinguishable from a pure
environmental noise and therefore the insect is unable to recognize
any signal of \emph{N. viridula} individuals. This implies that no
significant response is observed in terms of percentage of
source-direction movements (SDMs $\sim 50\%$).

\subsection{Threshold Stochastic Resonance} \label{sec:4} The
presence of a maximum in the behavior of SDM percentage as a
function of $D$ can be explained either by the threshold phenomenon,
or non-dynamical, stochastic resonance (TSR).

\begin{figure}
\begin{center}

\resizebox{0.7\columnwidth}{!}{%
\includegraphics{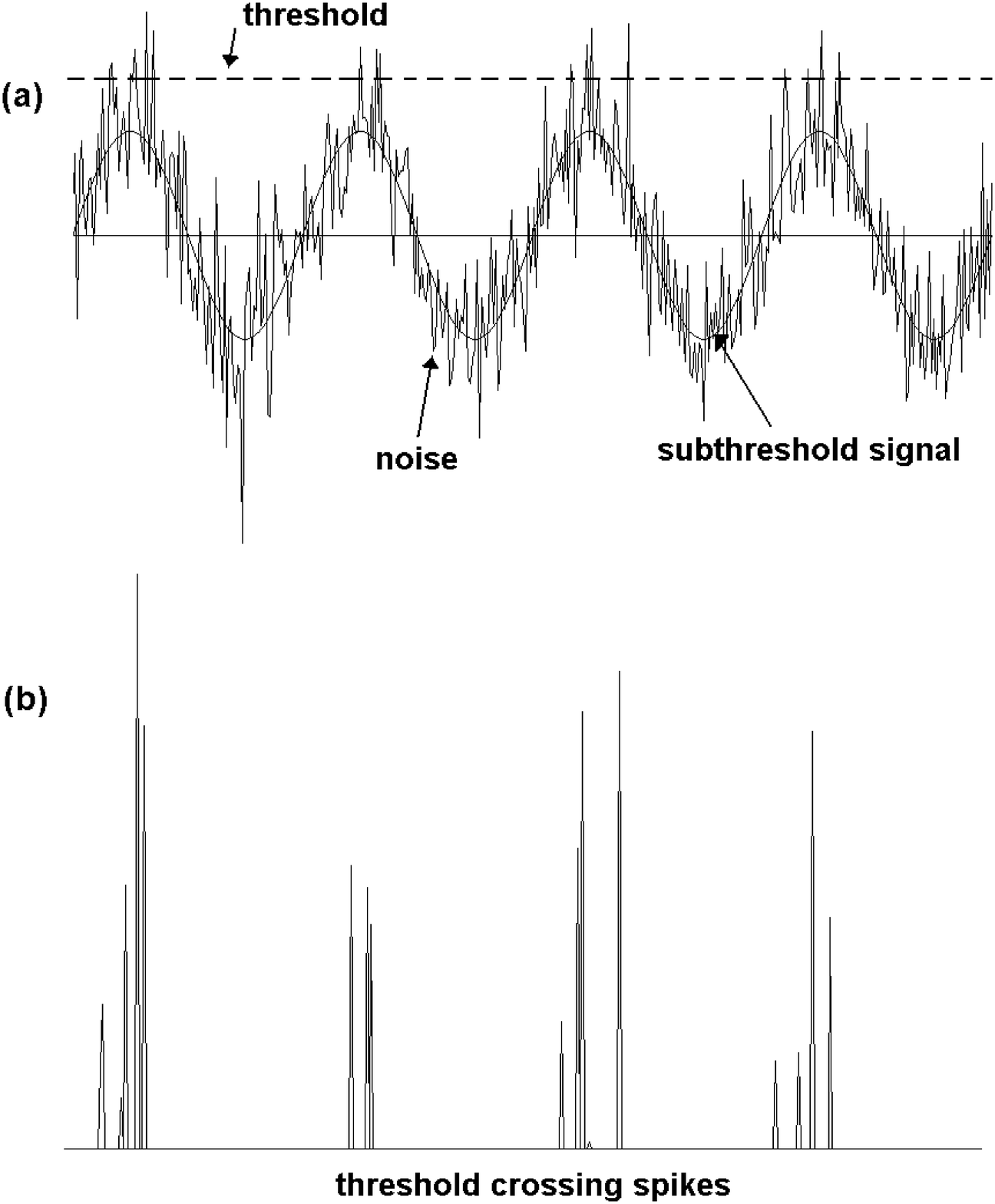}
}
\caption{Evolution driven by a sinusoidal function plus noise. (a)
Time series generated by consecutive pulses (dashed line: threshold
level, solid line: mean value of the periodic signal); (b) Temporal
sequence of threshold crossing events.}
\label{fig:3}       
\end{center}
\end{figure}
%
\begin{figure*}[h]
\begin{center}
\resizebox{1.0\columnwidth}{!} {%
\includegraphics{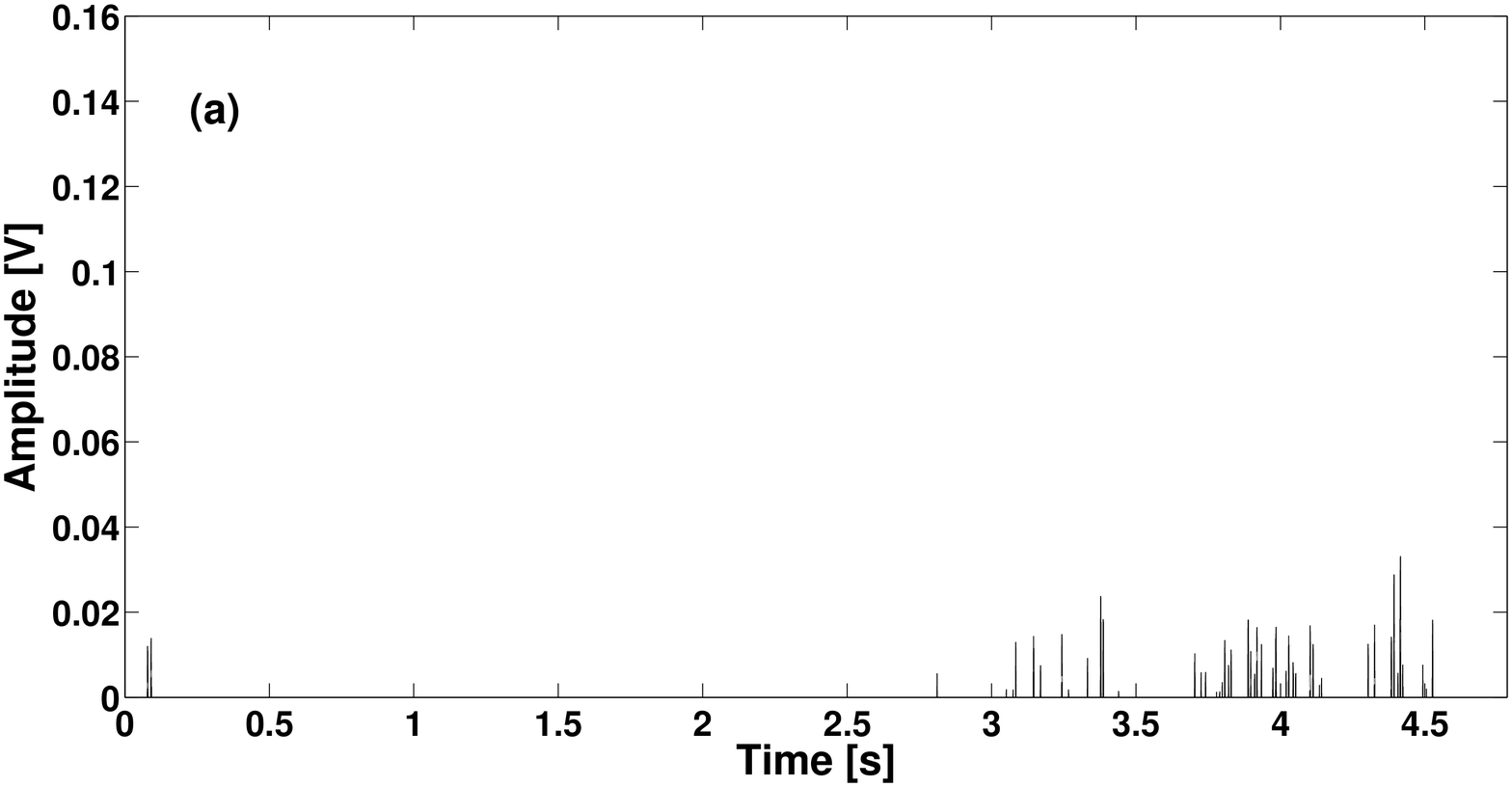}
\includegraphics{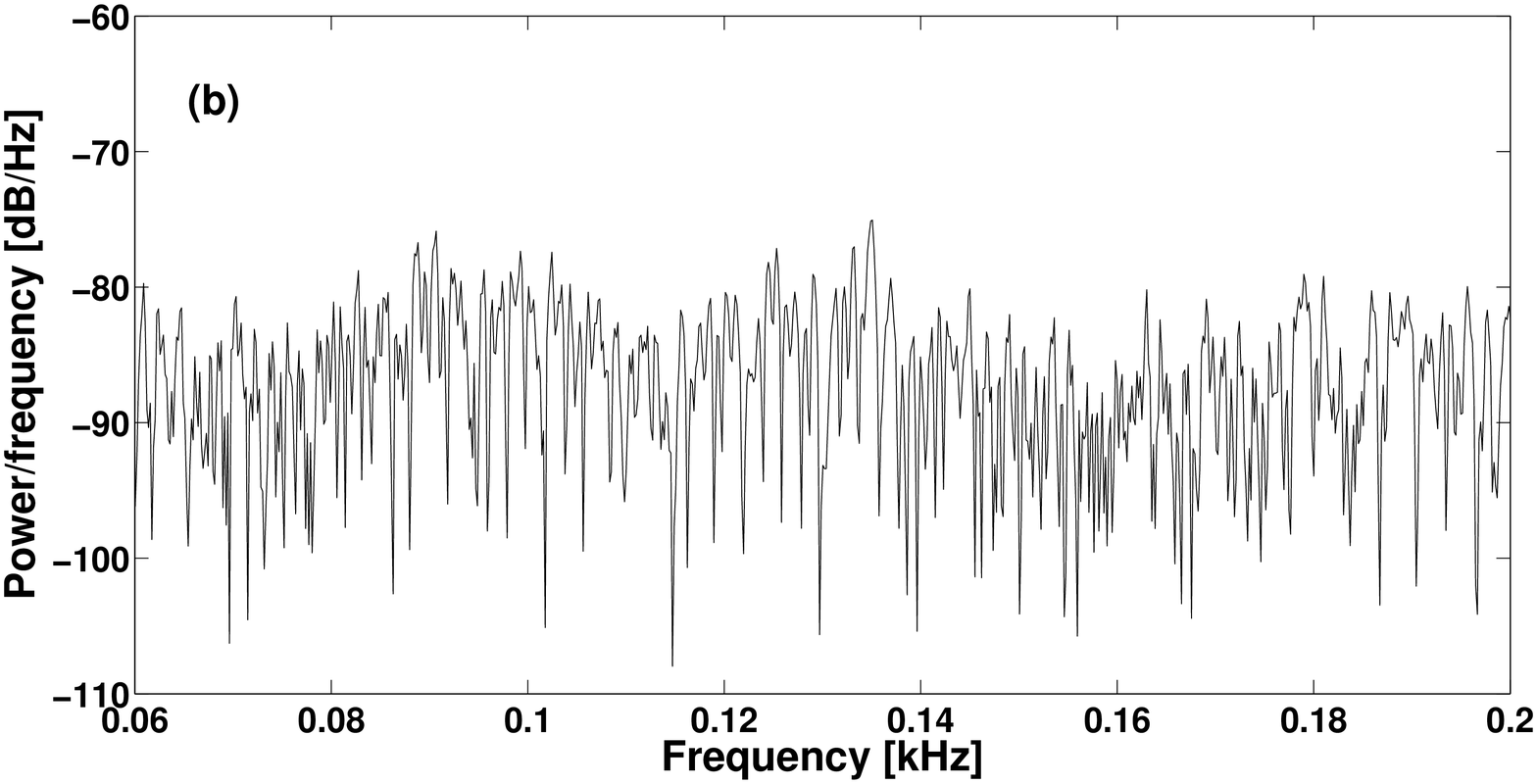}
}
\resizebox{1.0\columnwidth}{!} {%
\includegraphics{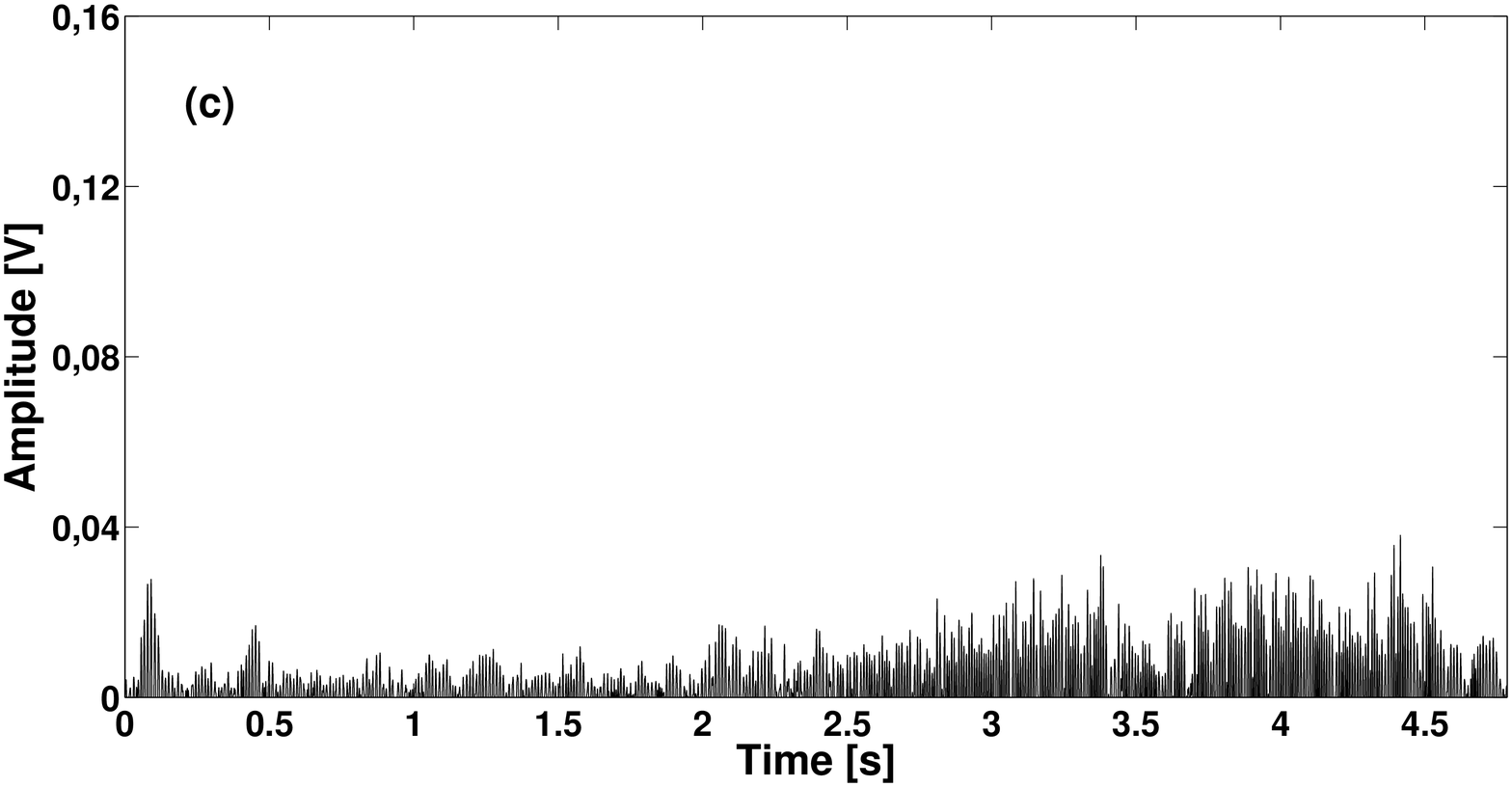}
\includegraphics{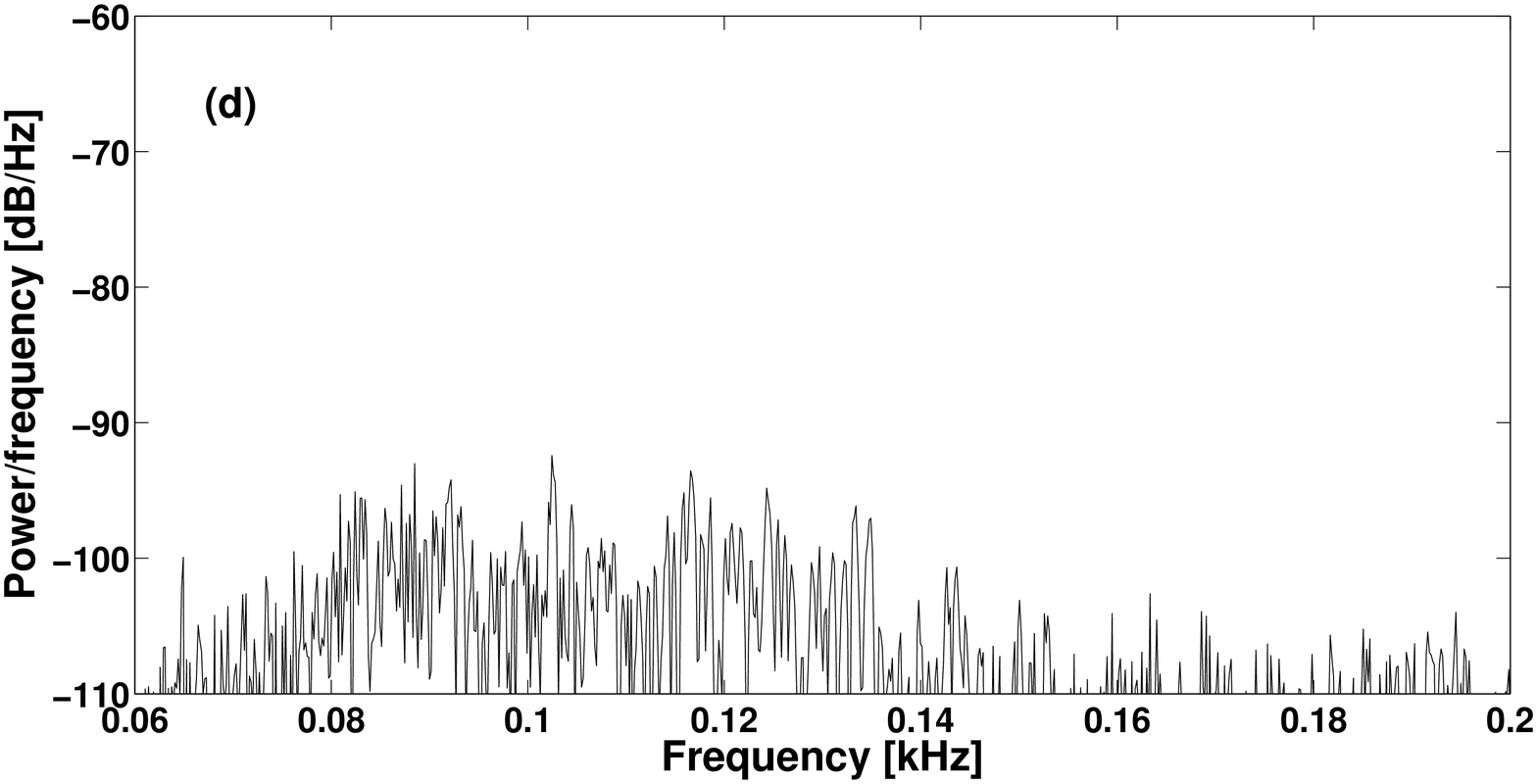}
}
\resizebox{1.0\columnwidth}{!} {%
\includegraphics{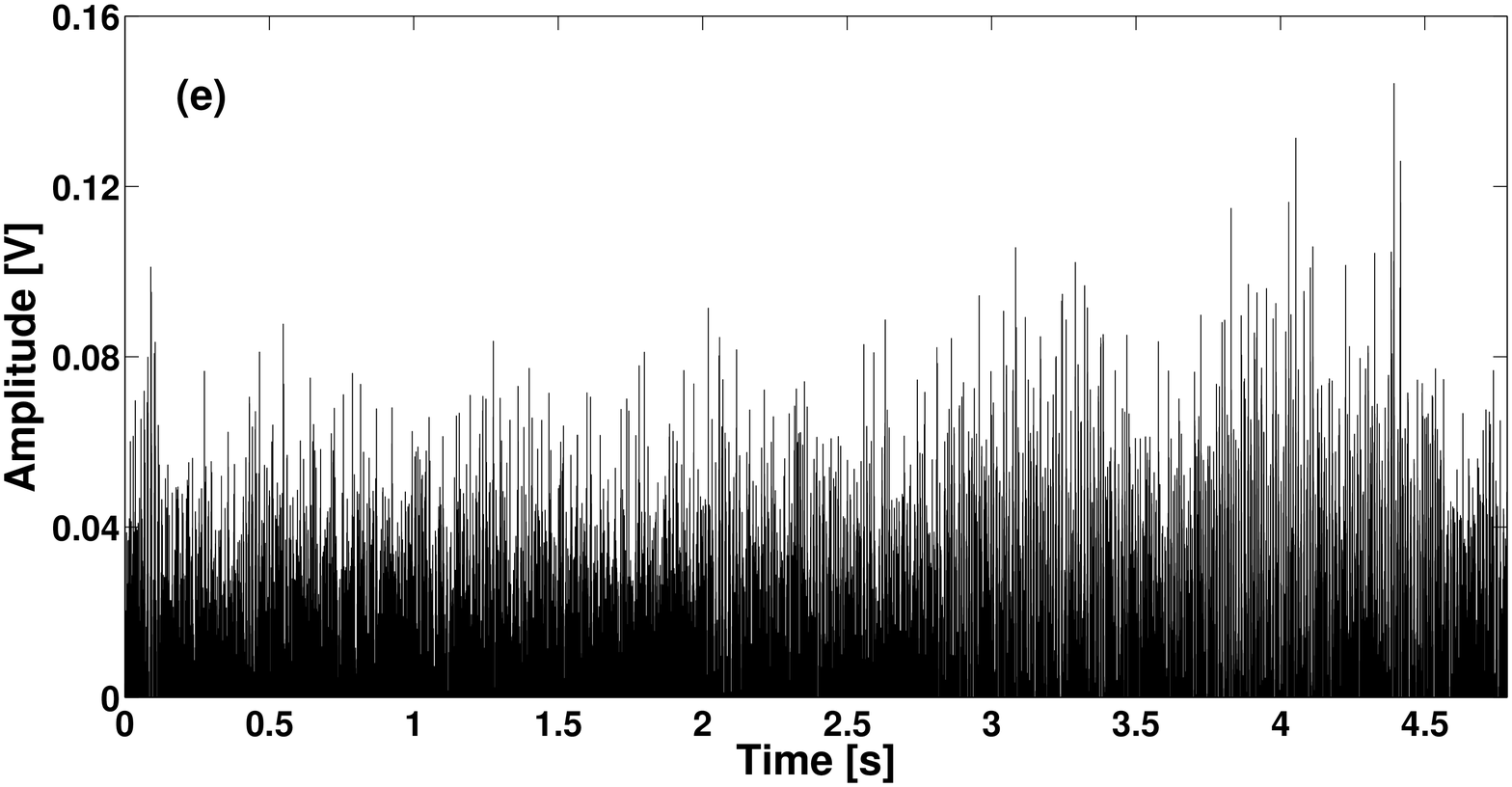}
\includegraphics{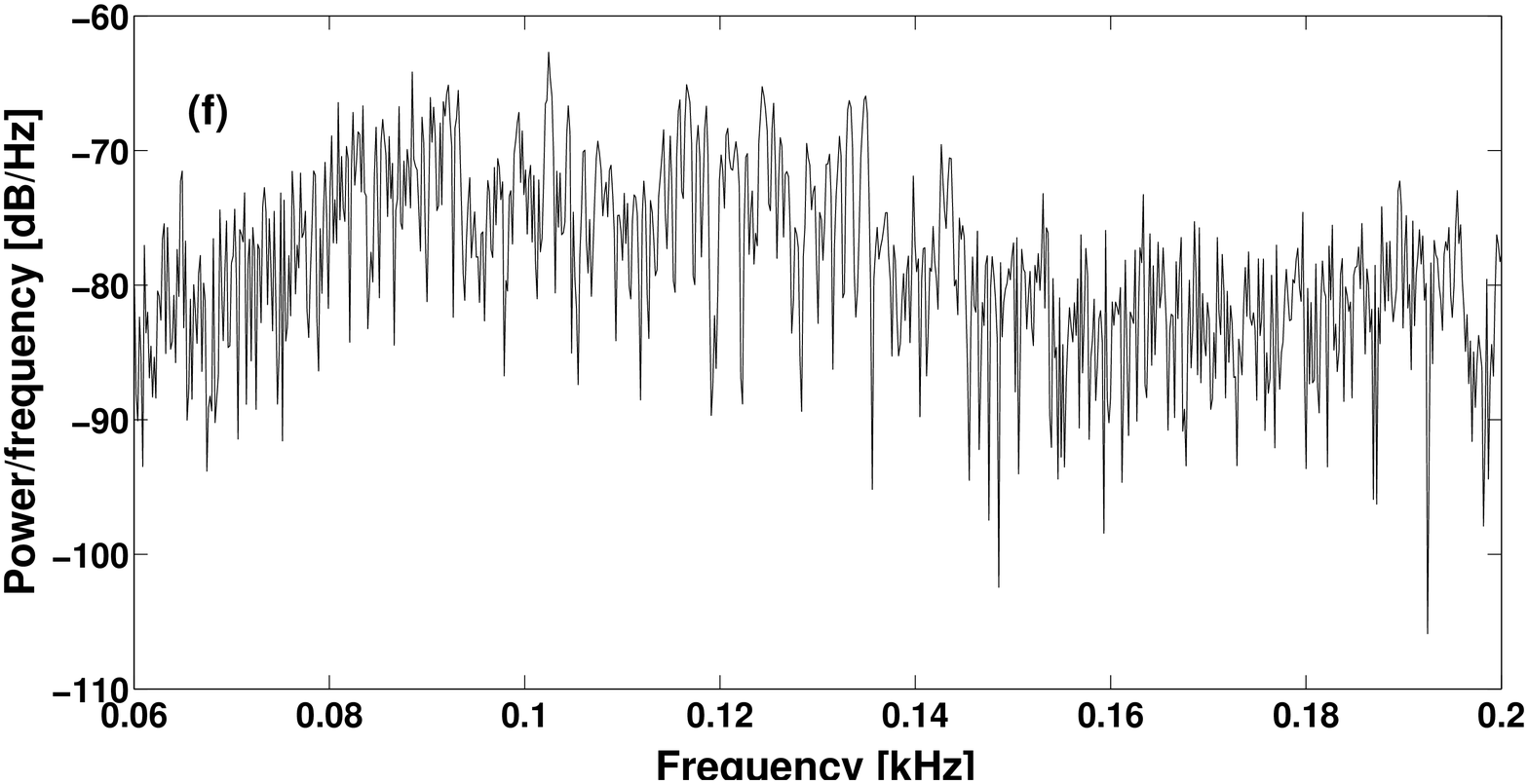}
}
\vspace*{0.7cm}       
\caption{Temporal evolution of the simulated calling signal over the
threshold for noise intensity $D = 2.6 \cdot 10^{-6}$ $V^2$ (a), $D
= 1.30 \cdot 10^{-5}$ $V^2$ (b) and $D = 1.0 \cdot 10^{-3}$ $V^2$
(c). The corresponding power spectral densities are shown in panels
b, d, f. The threshold level is $s_{th} = 0.045$ V and RMS amplitude
of the subthreshold signal is $0.031$ $V^2$. In the figures (a), (c)
and (e) we have rescaled the values of the signal amplitude in such
a way that the zero value corresponds to the threshold value. }

\label{fig:4}       
\end{center}
\end{figure*}

Stochastic resonance (SR), initially observed in the temperature
cycles of the Earth \cite{Benzi}, is a counterintuitive phenomenon
occurring in a large variety of non-linear systems, whereby the
addition of noise to a weak periodic signal causes it to become
detectable or enhances the amount of transmitted information through
the
system~\cite{Mos94,Gin95,Pei95,Noz98,Lon98,Sto00,Wie94,Gam95,Wan00,Lin04,Vil98,Dou93,Rus99,Fre02,Gre00,Bul91,Nei02,Bah02,Gam98,Man94}.
When SR occurs, the response of the system undergoes resonance-like
behavior as a function of the noise level. In spite of the fact that
initially this phenomenon was restricted to bistable systems, it is
well known that SR appears in monostable, excitable, and
non-dynamical systems.

Here we report on experiments conducted on the response of \emph{N.
viridula} individuals to sub-threshold signals. The non-monotonic
behavior of SDM, as a function of the noise intensity (see
Fig.~\ref{fig:2}b), can be considered the hallmark of the threshold
stochastic resonance (TSR). This phenomenon is well described by an
extremely simple system, shown in Fig~\ref{fig:3}, and characterized
by: (i) an energetic activation barrier (threshold); (ii) a weak
coherent input such as a periodic signal (sub-threshold signal);
(iii) a source of noise which is inherent to the system, or is added
externally to the deterministic input~\cite{Mos94,Gin95,Pei95}.
Since the three ingredients are often present in nature and the idea
of the existence of a threshold is quite intuitive, TSR has migrated
into many different fields, so that during the last decades a
considerable amount of literature on this subject has appeared in
several areas of science and engineering.
%
\begin{figure}
\begin{center}
%
\resizebox{0.7\columnwidth}{!}{%
\includegraphics{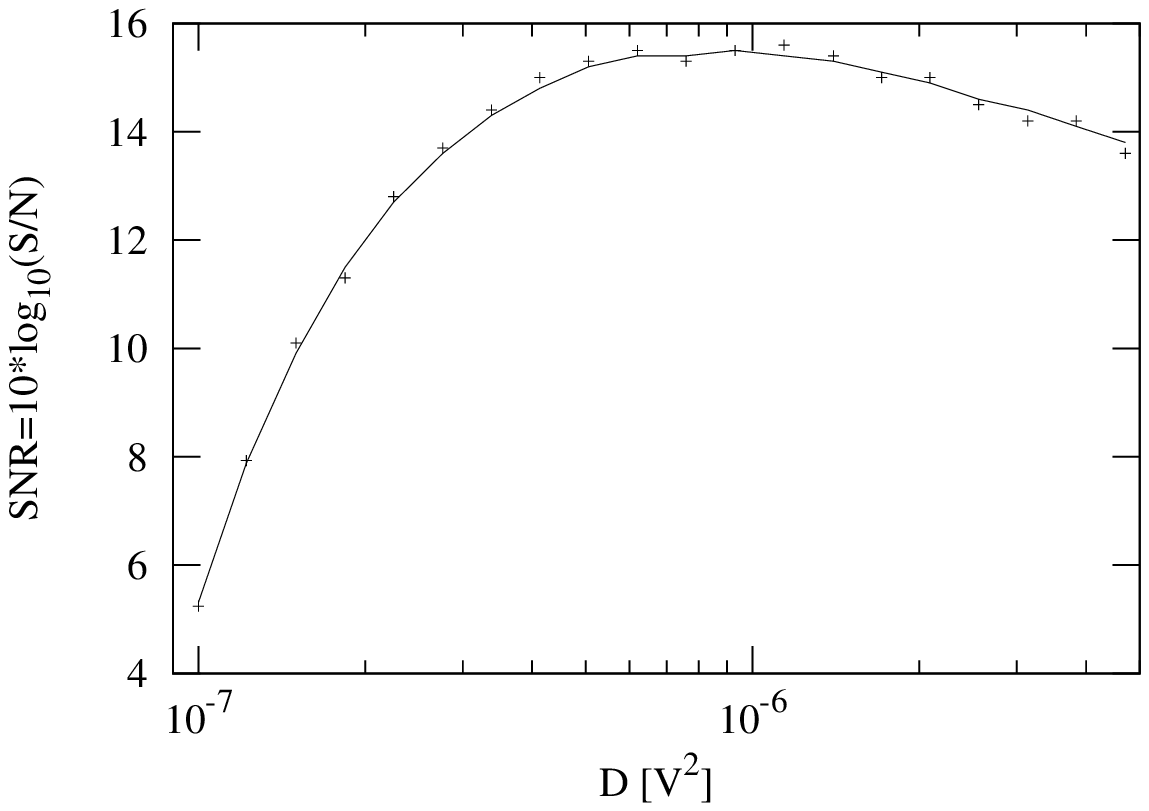}
}
\caption{Signal to noise ratio versus variance noise \emph{D} of the
output signal model when the input female calling song is
subthreshold, at the dominant frequency $f = 102.5$. All the other
parameters are the same of Fig.~\ref{fig:4}.}
\label{fig:5}       
\end{center}
\end{figure}
We have simulated a system with a threshold $0.045$ V
 and a subthreshold signal of RMS amplitude $0.031$ V 
(a. u.), obtained by the recorded female calling song. In
Fig.~\ref{fig:4} we show the output signal, and the corresponding
PSD, for three different levels of noise added to the subthreshold
deterministic signal (calling song). In the
Figs.~\ref{fig:4}a,~\ref{fig:4}c,~\ref{fig:4}e we have rescaled the
values of the signal amplitude in such a way that the zero value
corresponds to the threshold value. For low noise intensities the
signal crosses the threshold (dashed line in Fig.~\ref{fig:4}a) very
rarely, and in the corresponding PSD (Fig.~\ref{fig:4}b) no
frequency shows any significant power enhancement. By increasing the
noise level the threshold crossings become more frequent
(Fig.~\ref{fig:4}c) and the PSD appears to take a larger value for
$f = 102.5$ Hz (Fig.~\ref{fig:4}d), that is the dominant frequency
contained in the input signal. A further increase of noise intensity
produces a degradation of the signal, a loss of coherence in the
temporal sequence (Fig.~\ref{fig:4}e) and a reduction of the main
peak ($f = 102.5$ Hz) in the PSD (Fig.~\ref{fig:4}f). The
signal-to-noise ratio (SNR) at $f = 102.5$ Hz is reported in
Fig.~\ref{fig:5}. For each value of the noise intensity we have
performed $5000$ numerical realizations. The noise intensity for
which the SNR is maximum is $D \approx 1.17 \cdot 10^{-5}$, which is
very near the value of the noise intensity that maximizes the
percentage of SDMs (see Fig.~\ref{fig:2}b). The results obtained
from our model suggest that in the biological system analyzed,
stochastic resonance plays a key role, since it permits information
to be extracted from a weak deterministic signal, thanks to the
constructive action of environmental noise. In other words there is
a suitable noise intensity which maximizes the activating behavior
of the green bugs and this effect can be described by the simplest
threshold model which shows stochastic resonance. In
Fig.~\ref{fig:5} we report also the best fitting curve of the
simulations (cross points in the figure) obtained by the formula of
the SNR for a single frequency coherent signal~\cite{Mos94}
\begin{equation}
SNR=c\thinspace \log{[\frac{a}{D^{2}}\thinspace
\exp{(-\frac{b}{D})}]},
\end{equation}
where $a = 6.6 \cdot 10^{-5}$, $b = 1.70 \cdot 10^{-6}$, and $c =
2.18$.

\section{The Polymer Chain Model and MD Simulations}\label{mod} 
\begin{figure}[htbp]
\begin{center}
\includegraphics [width=8cm,height=5cm]{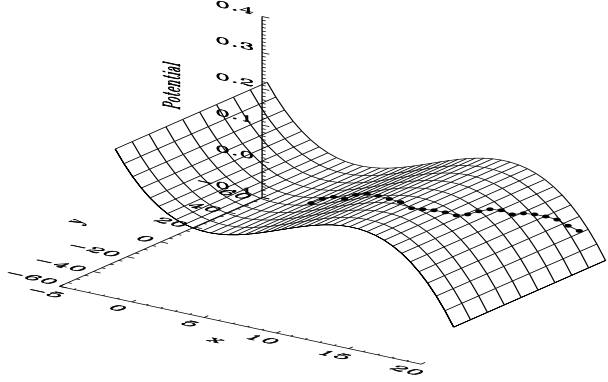}
\caption{3D-view of the potential energy
$U_{\rm Ext}$, which is included in our system to simulate the
presence of a barrier to be surmounted by the polymer. A sketch of
the translocating chain molecule is shown.} \label{fig1}
\end{center}
\end{figure}

The
polymer is modeled by a semi-flexible linear chain of $N$ beads
connected by harmonic springs \cite{Rou53}. Both excluded volume
effect and van der Waals interactions between all beads are kept
into account by introducing a Lennard-Jones (LJ) potential. In order
to confer a suitable stiffness to the chain, a bending recoil torque
is included in the model, with a rest angle $\theta_0=0$ between two
consecutive bonds. The total potential energy of the modeled chain
molecule is $U=U_{\rm Har}+U_{\rm Bend}+U_{\rm LJ}$ with
\begin{eqnarray}
&\displaystyle U_{\rm Har}=\sum_{i=1}^{N-1} K_{\rm r}(r_{i,i+1}-d)^2 \label{eq1}\\
&\displaystyle U_{\rm Bend}=\sum_{i=2}^{N-1} K_{\rm
\theta}(\theta_{i-1,i+1}-\theta_0)^2 \label{eq2}\\
&\displaystyle U_{\rm LJ}=4\epsilon_{\rm LJ}\sum_{i,j (i\neq
j)}\left[\left(\frac{\sigma}{r_{ij}}\right)^{12}-\left(\frac{\sigma}{r_{ij}}\right)^6\right]
\label{eq3}
\end{eqnarray}
\noindent where $K_{\rm r}$ is the elastic constant, $r_{ij}$ the
distance between particles $i$ and $j$, $d$ the equilibrium distance
between adjacent monomers, $K_{\rm \theta}$ the bending modulus,
$\epsilon_{\rm LJ}$ the LJ energy depth and $\sigma$ the monomer
diameter. The effect of temperature fluctuations on the dynamics of
a chain polymer escaping from a metastable state is studied in a
two-dimensional environment. The polymer translocation is modeled as
a stochastic process of diffusion in the presence of a potential
barrier having the form:
\begin{eqnarray}
&U_{\rm Ext}(x)=ax^2-bx^3 \label{eq4}
\end{eqnarray}
\noindent with parameters $a=3\cdot10^{-3}$ and $b=2\cdot10^{-4}$,
as already adopted in Ref.~\cite{Piz09}. A three-dimensional view of
$U_{\rm Ext}$ is plotted in Fig.~\ref{fig1}.
The drift of the $i^{\rm th}$ monomer of the chain molecule is
described by the following overdamped Langevin equations:
\begin{eqnarray}
\frac{dx_i}{dt}&=&-\frac{\partial{U_{ij}}}{\partial{x}}-\frac{\partial{U_{\rm
Ext}}}{\partial{x}}+\sqrt{D}\xi_{\rm x}+A\cos(\omega t+\phi) \label{eq5}\\
\frac{dy_i}{dt}&=&-\frac{\partial{U_{ij}}}{\partial{y}}+\sqrt{D}\xi_{\rm
y} \label{eq6}
\end{eqnarray}
\noindent where $U_{ij}$ is the interaction potential between the
$i^{\rm th}$ and $j^{\rm th}$ beads, $\xi_x$ and $\xi_y$ are white
Gaussian noise modeling the temperature fluctuations, with the usual
statistical properties, namely $\langle\xi_k(t)\rangle=0$ and
$\langle\xi_k(t)\xi_l(t+\tau)\rangle=\delta_{(k,l)}\delta(\tau)$ for
$(k,l=x,y)$. $A$ and $\omega$ are respectively the amplitude and the
angular frequency of the forcing field and $\phi$ a randomly chosen
initial phase. In our simulations, the time $t$ is scaled with the
friction parameter $\gamma$ as $t=t_r/\gamma$, where $t_r$ is the
real time of the process. The standard Lennard-Jones time scale is
$\tau_{\rm LJ}=(m\sigma^2/\epsilon_{\rm LJ})^{1/2}$, where $m$ is
the mass of the monomer. A bead of a single-stranded DNA is formed
approximately by three nucleotide bases and then $\sigma\sim 1.5$ nm
and $m\approx 936$ amu \cite{Luo08}. Orders of magnitude of the
quantities involved in the process are nanometers for the
characteristic lengths of the system (polymer and barrier extension)
and microseconds for the time domain. A set of $10^3$ numerical
simulations has been performed for different values of the frequency
of the forcing field and two values of the noise intensity $D$,
namely $D=1.0, 4.0$. The values of the potential energy parameters
are: $K_{\rm r}=K_{\rm \theta}=10$, $\epsilon_{\rm LJ}=0.1$,
$\sigma=3$ and $d=5$, in arbitrary units (AU). The amplitude of the
electric forcing field is $A=5\cdot10^{-2}$ in AU, because it is
scaled with $\gamma$. The number of monomers $N$ is 20. The initial
spatial distribution of the polymer is with all monomers at the same
coordinate $x_0=0$, corresponding to the local minimum of the
potential energy of the barrier. Every simulation stops when the $x$
coordinate of the center of mass of the chain reaches the final
position at $x_f=15$.
\begin{figure}[htbp]
\begin{center}
\includegraphics [width=8cm,height=5cm]{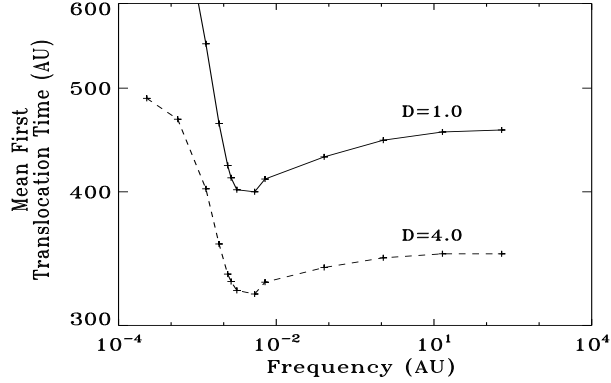}
\caption{MFTT vs.~frequency of the forcing field for two different
values of the noise intensity $D$. The values of the potential
energy parameters are: $K_{\rm r}=K_{\rm \theta}=10$, $\epsilon_{\rm
LJ}=0.1$, $\sigma=3$ and $d=5$, in arbitrary units (AU). The
amplitude of the electric forcing field is $A=5\cdot10^{-2}$ (AU).
The number of monomers $N$ is 20.}\label{fig2}
\end{center}
\end{figure}
\begin{figure}[htbp]
\begin{center}
\includegraphics [width=8cm,height=13cm]{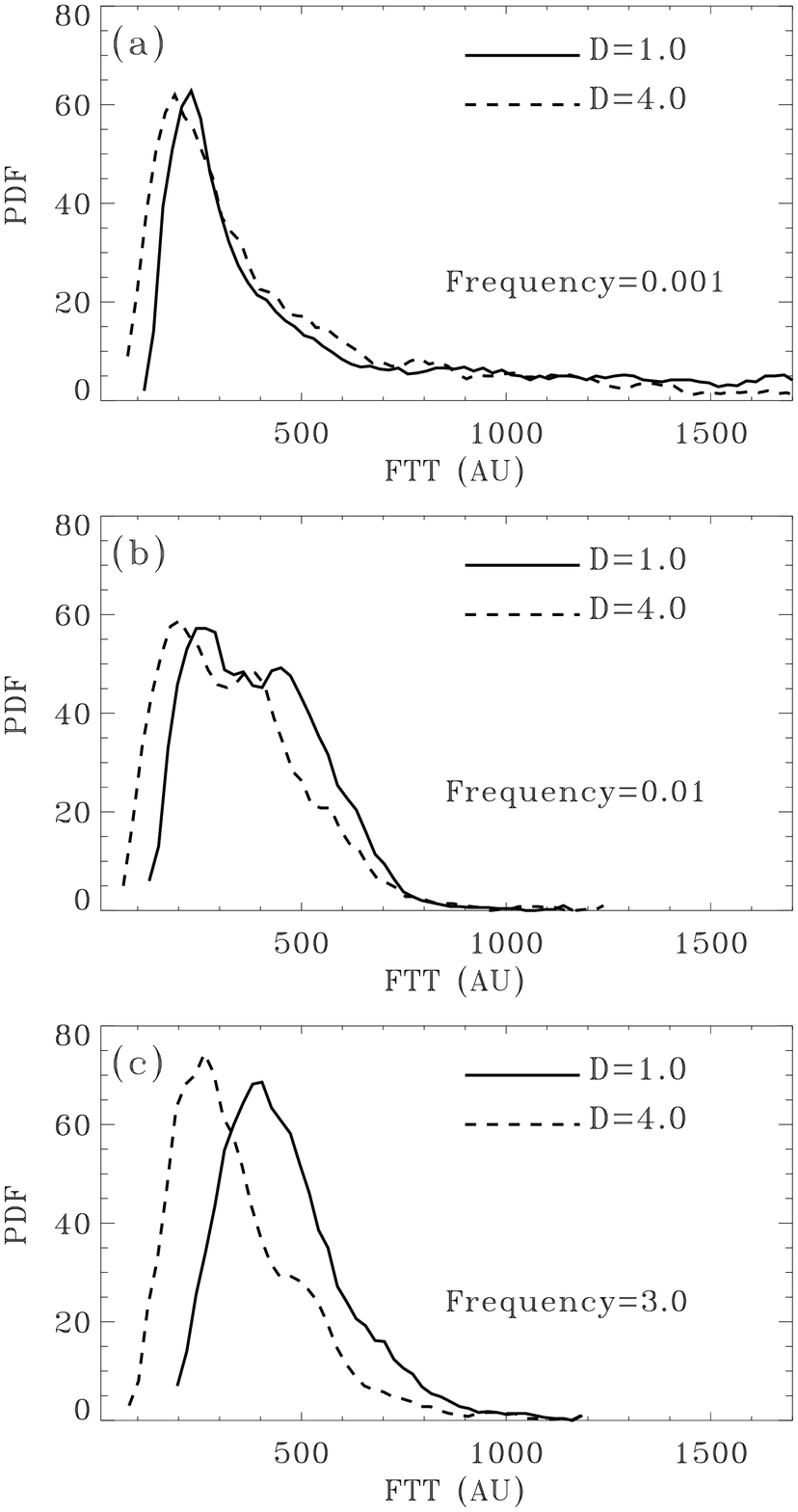}
\caption{Probability density function (PDF) of the first
translocation time (FTT). Each panel shows two PDFs, each one
characterized by a specific value of the noise intensity. The three
panels differs for the frequency of the forcing field. The panel (a)
shows the time distribution in the low frequency region. The long
tails indicate that the polymer crosses the
potential barrier with a longer mean time. In the panel (b) the FTTs
are distributed towards shorter values, because of the lowest time
scale characterizing the translocation process in the resonant
activation regime. The panel (c) shows the probability distribution
for the high frequency domain, where the time scale is the same that
is characterized by the presence of a static potential.}\label{fig3}
\end{center}
\end{figure}

\subsection{Results and discussion}\label{res} The MFTT shows three
different translocation regimes as a function of the frequency
(Fig.~\ref{fig2}). In the low frequency domain ($\omega<10^{-3}$),
the period of the forcing field oscillations is very long with
respect to the typical values of the mean crossing time of the chain
molecule. In this regime the MFTT is equal to the average of the
crossing times over upper and lower configurations of the barrier,
and the slowest process determines the value of the mean crossing
time. As a consequence, the MFTT increases and we observe long tails
in the probability density function (PDF) shown in Fig.~\ref{fig3}a.
In the high frequency domain ($\omega>10^{-1}$), a saturation of the
translocation time is obtained. In this case, very rapid
oscillations act on the polymer motions as the mean potential, i.~e.
the static field, and therefore the MFTT becomes equal to that
obtained without any additional periodic driving. In other words,
the polymer chain feels the average potential barrier. For
intermediate frequencies ($10^{-2}<\omega<10^{-1}$), the crossing
event is strongly correlated with the potential oscillations and the
MFTT vs.~$\omega$ exhibits a minimum at a resonant oscillation rate.
This frequency region corresponds to periods of oscillations which
are of the same order of magnitude of the mean time the polymer
takes to cross a static barrier with its shape corresponding to the
lowest configuration of the oscillating potential. In other words,
the potential remains around its lowest configuration for enough
time to allow the polymer to exit and, even in the case of an
initially high or intermediate value of the height of the barrier,
the potential feature turns into the lowest configuration within a
sufficiently short time lag to facilitate the translocation process.
The polymer, driven by a periodic field oscillating at a period
comparable with a characteristic time of the crossing dynamics,
reaches a resonant regime that accelerates the translocation
process. For each of the frequency values, the thermal noise
intensity $D$ is able to speed up or slow down the crossing process,
as described by the three frequency regions (Fig.~\ref{fig2}) and
the corresponding translocation dynamics \cite{Pizzolato2010}. The probability density
function of the first translocation time (FTT) is shown in
Fig.~\ref{fig3} for three frequency values characterizing the
different dynamical domains. Each panel shows two PDFs, each one
characterized by a specific value of the noise intensity. In the
resonant activation regime (Fig.~\ref{fig3}b) the PDFs do not
present the long tail at higher crossing times, observed in
Fig.~\ref{fig3}a. Consequently, the MFTT reduces its value. The PDF
assumes an interesting two-peaks structure that suggests the
presence of two characteristic times of translocation. This feature,
being present both at low and high noise intensity, can be ascribed
to two different translocation dynamics of the polymer chain
surmounting the barrier. In the high frequency domain
(Fig.~\ref{fig3}c) the PDFs show the characteristic feature of the
static potential case.

\section{Verhulst Model with L\'{e}vy White Noise Excitation}

In considering how the population density $x\left(t\right)$ may
change with time $t$, Verhulst proposed the following equation
\begin{equation}
\frac{dx}{dt}=rx\left( 1-\frac{x}{\Omega }\right) . \label{F-1}
\end{equation}
where there is the Malthus term with the rate constant $r$ and a
saturation term with the $\Omega$ factor, which is the upper limit
for the population growth due to the availability of the resources.

Really the parameters $r$ and $\Omega$ are not constant. In fact the
parameter $r$ changes randomly due to season fluctuations, and the
parameter $\Omega$ fluctuates due to the environmental interaction
which causes the random availability of resources. As a consequence
we have the following stochastic Verhulst equation
\begin{equation}
\frac{dx}{dt}=r\left( t\right) x\left[ 1-\frac{x}{\Omega
\left(t\right) }\right] . \label{F-2}
\end{equation}
In the context of macromolecular self-replication, the model
equation~(\ref{F-2}), with constant $\Omega$ and a white Gaussian
noise in $r\left( t\right)$, was numerically studied in
Ref.~\cite{Leu88} and the critical slowing down, i.e. a divergence
of the relaxation time at some noise intensity, was found. Later
Jackson and co-authors~\cite{Jac89} investigated the same model, by
analog experiment and digital simulations. They analyzed
specifically in detail the nonlinear relaxation time defined
as~\cite{Bin73}
\begin{equation}
T=\frac{\int\nolimits_{0}^{\infty}\left[\left\langle x\left(
t\right) \right\rangle -\left\langle x\left(\infty\right)
\right\rangle \right] dt}{x\left(0\right) - \left\langle x\left(
\infty\right) \right\rangle} \label{F-3}
\end{equation}
and did not observe the critical slowing down. They explained this
discrepancy by the incorrect approximate truncation of the
asymptotic power series for $T$ used in Ref.~\cite{Leu88}. The
stability conditions were derived in Ref.~\cite{Gol03}. Similar
investigations for colored Gaussian noise $r(t)$ were performed in
Ref.~\cite{Man90}, where a monotonic dependence of the relaxation
time and the correlation time on the noise intensity was found. The
generalization of Eq.~(\ref{F-2}), to study a
Bernoulli-Malthus-Verhulst model driven by a multiplicative colored
noise, was analyzed recently in
Ref.~\cite{Cal07}.\\
\indent The evolution of the mean value in the case of
Eq.~(\ref{F-2}) with constant $r$ and white Gaussian noise
excitation $\beta\left( t\right) =r/\Omega \left( t\right)$ was
considered in Refs.~\cite{Ciu93,Suz82,Suz82a,Bre82,Mak85,Mor86}. In
Refs.~\cite{Mak85,Mor86} the authors, using perturbation technique,
obtained the exact expansion in power series on noise intensity of
all the moments and found the long-time decay of $t^{-1/2}$. In
Ref.~\cite{Ciu93} the authors derived the long-time behavior of all
the moments of the population density by means of an exact
asymptotic expansion of the time averaged process generating
function, and found the same asymptotic behavior of $t^{-1/2}$ at
the critical point. This very slow relaxation of the moments near
the critical point is the phenomenon of critical slowing down.\\
\indent In the present chapter, using the previously obtained
results for a generalized Langevin equation with a L\'{e}vy noise
source~\cite{Dub05,Dub08}, we investigate the transient dynamics of
the stochastic Verhulst model with a fluctuating growth rate and a
constant value for the saturation population density $\Omega$, that
is $\Omega=1$. The exact results for the mean value of the
population density and its nonstationary probability distribution
for different types of white non-Gaussian excitation $r\left(
t\right)$ are obtained. We find the interesting noise-induced
transitions for the probability distribution of the population
density and the relaxation dynamics of its mean value for Cauchy
stable noise. Finally we obtain a nonmonotonic behavior of the
nonlinear relaxation time as a function of the Cauchy noise
intensity.

\section{Stochastic Verhulst Equation with Non-Gaussian Fluctuations of Growth
Rate}

Let us consider Eq.~(\ref{F-2}) with a constant saturation value
$\Omega=1$, namely
\begin{equation}
\frac{dx}{dt}=r\left(  t\right)  x\left(  1-x\right)  . \label{F-4}
\end{equation}
After changing variable $y=\ln[x/(1-x)]$, we obtain
\[
y\left(t\right) = y\left(0\right) +\int\nolimits_{0}^{t}r\left(
\tau\right) d\tau \,
\]
and the exact solution of Eq.~(\ref{F-4}) is
\begin{equation}
x\left(t\right) = \left(1+\frac{1-x_{0}}{x_{0}}\,\exp\left\{ -\int
\nolimits_{0}^{t}r\left(\tau\right) d\tau\right\} \right)^{-1},
\label{F-5}
\end{equation}
where $x_{0}=x\left(0\right)$. Now by substituting in
Eq.~(\ref{F-5}) the following expression for the random rate $r(t)$
\begin{equation}
r\left( t\right) =r+\xi\left( t\right), \label{F-6}
\end{equation}
where $r>0$ and $\xi\left( t\right)$ is an arbitrary white
non-Gaussian noise with zero mean, we can rewrite the
solution~(\ref{F-5}) as
\begin{equation}
x\left(t\right) =
\left(1+\frac{1-x_{0}}{x_{0}}\,e^{-rt-L\left(t\right)}\right)^{-1}.
\label{F-7}
\end{equation}
Here $L\left(t\right)$ denotes the so-called L\'{e}vy random process
with $L\left(0\right)=0$, and
$\xi\left(t\right)=\dot{L}\left(t\right) $. As it was shown in
Refs.~\cite{Dub05,Dub08,Fel71}, L\'{e}vy processes having stationary
and statistically independent increments on non-overlapping time
intervals belongs to the class of stochastic processes with
infinitely divisible distributions. As a consequence, the
characteristic function of $L\left( t\right)$ can be represented in
the following form (see Eq.~(6) in ~\cite{Dub05})
\begin{equation}
\left\langle e^{iuL\left(  t\right)  }\right\rangle =\exp\left\{
t\int\nolimits_{-\infty}^{+\infty}\frac{e^{iuz}-1-iu\sin
z}{z^{2}}\,\rho\left( z\right)  dz\right\}  , \label{F-8}
\end{equation}
where $\rho\left(z\right)$ is some non-negative kernel function. The
case $\rho\left(z\right)=2D\delta\left(  z\right)$ corresponds to a
white Gaussian noise excitation $\xi\left(t\right)$, while for a
symmetric L\'{e}vy stable noise $\xi\left( t\right)$ with index
$\alpha$ we have a power-law kernel $\rho\left( z\right)=Q\left\vert
z\right\vert ^{1-\alpha}$, with $0<\alpha<2$.

\begin{figure}[ptbh]
\centering{\resizebox{7cm}{!}{\includegraphics{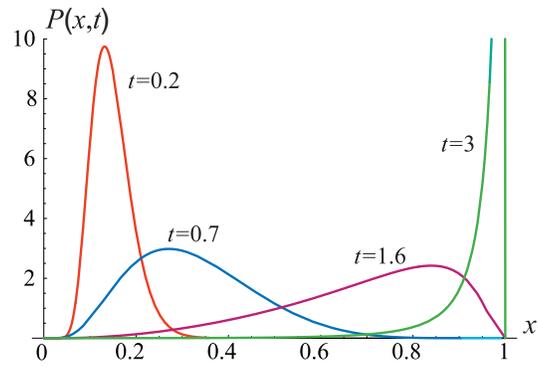}}}
\caption{Time evolution of the probability distribution of the
population density for white Gaussian noise excitation with
intensity $D$. The values of the parameters are: $x_0=0.1$, $r=2$,
$D=0.3$.} \label{fig-1}
\end{figure}

In the model under consideration the stationary probability
distribution has; (i) a singularity at the stable point $x=1$ for
white Gaussian noise; and (ii) two singularities at both stable
points $x=0$ and $x=1$ for L\'{e}vy noise. To analyze the time
behavior of the probability distribution in the transient dynamics
it is better not to use the Kolmogorov equation for the probability
density $P\left(x,t\right)$, but rather the exact
solution~(\ref{F-7}). Using the standard theorem of the probability
theory regarding a nonlinear transformation of a random variable, we
find from Eq.~(\ref{F-7})
\begin{equation}
P\left(x,t\right) =\frac{1}{x\left(1-x\right)}P_{L}\left(\ln
\left[\frac{\left(1-x_{0}\right)x}{x_{0}\left(1-x\right)}\right]
-rt,t\right), \label{F-9}
\end{equation}
where $P_{L}\left(z,t\right)$ is the probability density
corresponding to the characteristic function~(\ref{F-8}). For a
white Gaussian noise $\xi\left(t\right)$, this distribution reads
\begin{equation}
P_{L}\left(  z,t\right)  =\frac{1}{2\sqrt{\pi Dt}}\exp\left\{
-\frac{z^{2} }{4Dt}\right\}  . \label{F-10}
\end{equation}
The time evolution of the probability distribution
$P\left(x,t\right) $ for $D=0.3$, $r=2$, and $x_{0}=0.1$ is plotted
in Fig.~\ref{fig-1}.

As it is easily seen, the maximum of the unimodal distribution with
initial position at $x=0.1$ shifts with time towards the stable
point at $x=1$. At the same time, as it follows from
Eqs.~(\ref{F-9}) and~(\ref{F-10}), for all $t>0$ we have
\begin{equation}
\lim_{x\rightarrow 0^+}P\left(x,t\right) = \lim_{x\rightarrow
1^-}P\left(x,t\right) = 0. \label{F-11}
\end{equation}

\begin{figure}[ptbh]
\centering{\resizebox{7cm}{!}{\includegraphics{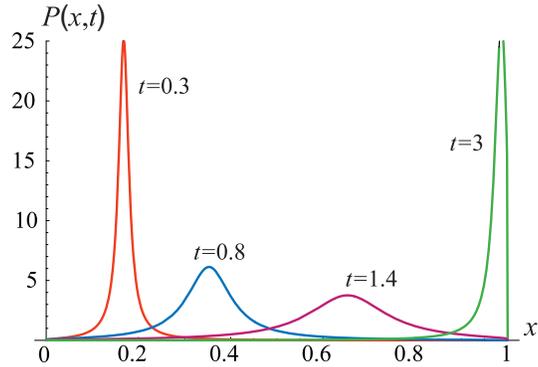}}}
\vskip-0.4cm\caption{Time evolution of the probability distribution
of the population density in the case of L\'{e}vy noise with
distribution~(\ref{F-12}). The values of the parameters are:
$x_0=0.1$, $r=2$, $K=0.2$.} \label{fig-2}
\end{figure}

The same picture is observed for another kernel function
$\rho\left(z\right) = Kz/\left(2\sinh z\right)$ $\left(K>0\right)$,
corresponding to a L\'{e}vy process $\eta\left(t\right)$ with finite
moments and the following probability density of increments
\begin{equation}
P_{L}\left(  z,t\right)
=\frac{2^{Kt-1}}{\pi^{2}\Gamma\left(Kt\right)} \Gamma\left(
\frac{Kt}{2}+\frac{iz}{\pi}\right) \Gamma\left(\frac{Kt}
{2}-\frac{iz}{\pi}\right)  , \label{F-12}
\end{equation}
where $\Gamma\left(x\right)$ is the Gamma function. The
corresponding time evolution of the probability distribution
$P\left( x,t\right)$ for $K=0.2$, $r=2$, and $x_{0}=0.1$ is shown in
Fig.~\ref{fig-2}.

\begin{figure}[ptbh]
\centering{\resizebox{7cm}{!}{\includegraphics{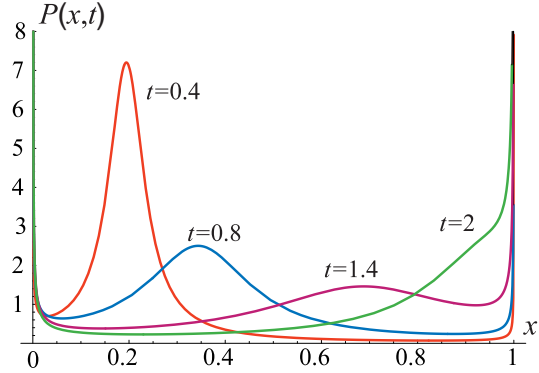}}}
\caption{Time evolution of the probability distribution of the
population density in the case of white Cauchy noise excitation. The
values of the parameters are: $x_0 = 0.1$, $r = 2$, $D_1 = 0.7$.}
\label{fig-3}
\end{figure}

\begin{figure}[ptbh]
\centering{\resizebox{7cm}{!}{\includegraphics{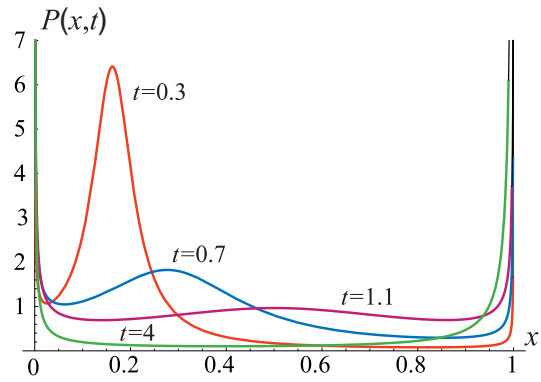}}}
\caption{Time evolution of the probability distribution of the
population density in the case of white Cauchy noise. The values of
the parameters are $x_0=0.1$, $r=2$, $D_1=1.2$.} \label{fig-4}
\end{figure}

\begin{figure}[ptbh]
\centering{\resizebox{7cm}{!}{\includegraphics{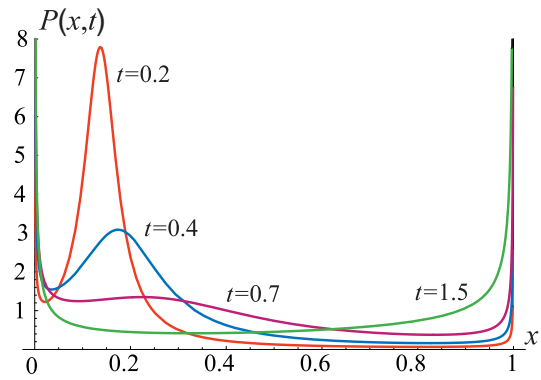}}}
\caption{Time evolution of the probability distribution of the
population density in the case of white Cauchy noise. The values of
the parameters are $x_0=0.1$, $r=2$, $D_1=1.7$.} \label{fig-5}
\end{figure}

A different situation we have for a Cauchy stable noise
$\xi\left(t\right)$ with constant kernel $\rho\left(z\right) = Q$
$\left( \alpha=1\right)$. After evaluation of the integral in
Eq.~(\ref{F-8}), the probability density of the L\'{e}vy process
increments takes the form of the well-known Cauchy
distribution~\cite{Fel71}
\begin{equation}
P_{L}\left(z,t\right) = \frac{D_{1}t}{\pi\left[z^{2}+
\left(D_{1}t\right)^{2}\right]}, \label{F-13}
\end{equation}
where $D_{1}=\pi Q$ is the noise intensity parameter. In such a case
from Eqs.~(\ref{F-9}) and~(\ref{F-13}) for all $t>0$ we find
\begin{equation}
\lim_{x\rightarrow 0^+}P\left(  x,t\right) =\lim_{x\rightarrow
1^-}P\left( x,t\right)  = \infty. \label{F-14}
\end{equation}
As a result, from an initial delta function we immediately obtain a
trimodal distribution for $t>0$ and then after some transition time
$t_c$ a bimodal one with two singularities at the stable points
$x=0$ and $x=1$ (see Figs.~\ref{fig-3}--\ref{fig-5}).
We should note that the transition from trimodal to bimodal
distribution is a general feature of the model in the presence of a
Cauchy stable noise, and it is not limited to some range of
parameters. In fact, from Eq.~(\ref{F-14}) and a delta function
initial distribution inside the interval $(0,1)$, this transition
always takes place.

In the following Figs.~\ref{fig-4} and~\ref{fig-5} we show the time
evolution of the probability distribution of the population density
for two other values of the noise intensity, namely $D_1 = 1.2$ and
$D_1 = 1.7$. As the noise intensity increases the probability
distribution shows two singularities near $x = 0$ and $x = 1$ with
different amplitude.

This transition in the shape of the probability distribution of the
population density is due to both the multiplicative noise and the
L\'{e}vy noise source. Using Eqs.~(\ref{F-9}) and~(\ref{F-13}) and
equating to zero the derivative of $P(x,t)$ with respect to $x$, we
obtain the following condition for the extrema in the range $0<x<1$,
and particularly for a minimum in the same interval
\begin{equation}
\frac{z(x,t)}{z(x,t)^2 + (D_1 t)^2} = x - \frac{1}{2} \; ,
\label{F-15}
\end{equation}
with
\begin{equation}
z(x,t) = \ln
\left[\frac{\left(1-x_{0}\right)x}{x_{0}\left(1-x\right)}\right] -rt
\,. \label{F-16}
\end{equation}
This condition can be solved graphically by finding the intersection
between the functions $y_1 = z(x,t)/(z(x,t)^2 + (D_1 t)^2)$ and $y_2
= x - 1/2$. This is done in the following
Figs.~\ref{fig-6}--\ref{fig-8}, where the function $y_1$ is plotted
for three different values of time and noise intensity. In each
figure the black blue curve (color on line) corresponds to the
critical value of time $t_c$ for which we have a noise induced
transition of the probability distribution of the population density
from trimodal to bimodal, that is from two minima and one maximum to
one minimum inside the interval $0 < x < 1$. The appearance of one
minimum in the probability distribution is the signature of this
transition.
\begin{figure}[ptbh]
\centering{\resizebox{7cm}{!}{\includegraphics{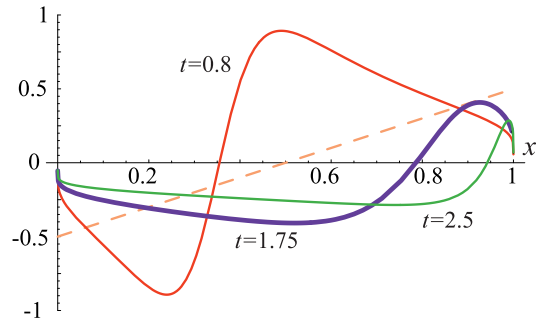}}}
\caption{Plots of both sides of Eq.~(\ref{F-15}) (white Cauchy
noise): function $y_1$ (solid curves), function $y_2$ (dashed
curve), for three values of time, namely: $t = 0.8,\, 1.75,\, 2.5$.
The critical time is $t_c = 1.75$ (black blue curve). The values of
the other parameters are: $x_0=0.1$, $r=2$, $D_1=0.7$.}
\label{fig-6}
\end{figure}
\begin{figure}[ptbh] \vspace{5mm}
\centering{\resizebox{7cm}{!}{\includegraphics{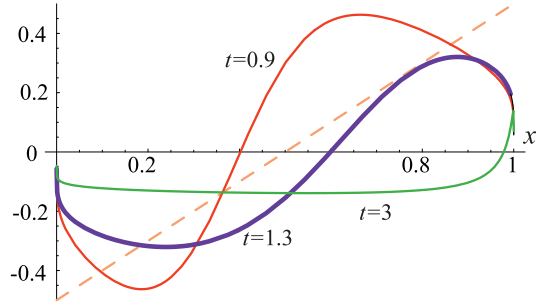}}}
\caption{Plots of both sides of Eq.~(\ref{F-15}) (white Cauchy
noise): function $y_1$ (solid curves), function $y_2$ (dashed
curve), for three values of time, namely: $t = 0.9,\, 1.3,\, 3$. The
critical time is $t_c = 1.3$ (black blue curve). The values of the
other parameters are: $x_0=0.1$, $r=2$, $D_1=1.2$.} \label{fig-7}
\end{figure}

The three values of the critical time $t_c$ corresponding to the
three values of the L\'{e}vy noise intensity investigated are: $D_1
= 0.7,~t_c = 1.75;~D_2 = 1.2,~t_c = 1.3;~D_3 = 1.7,~t_c = 0.95$. One
rough evaluation of the critical time $t_c$ is obtained by putting
equal to $1$ the scale parameter of the Cauchy distribution of
Eq.~(\ref{F-13}), that is $\tau_{c}\sim1/D_{1}$. The critical time
$t_c$ is the time at which the maximum and one minimum of the
probability distribution (see Figs.~\ref{fig-3}--\ref{fig-5})
coalesce in one inflection point and in this point $x$ the function
$y_2 = x - 1/2$ becomes tangent at the function $y_1$ (see
Figs.~\ref{fig-6}--\ref{fig-8}). It is interesting to note that the
critical time $t_c$ decreases with the noise intensity $D_1$. This
is because by increasing the noise intensity, more quickly the
population density reaches the two points near the boundaries $x =
0$ and $x = 1$.
\begin{figure}[ptbh] \vspace{5mm}
\centering{\resizebox{7cm}{!}{\includegraphics{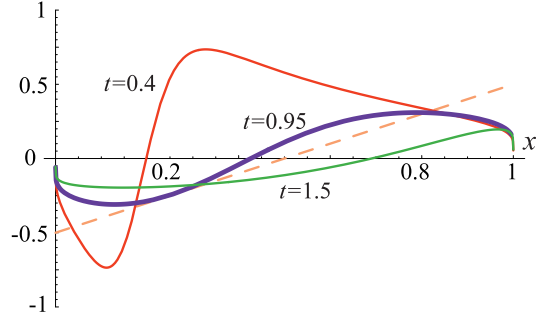}}}
\caption{Plots of both sides of Eq.~(\ref{F-15}) (white Cauchy
noise): function $y_1$ (solid curves), function $y_2$ (dashed
curve), for three values of time, namely: $t = 0.4,\, 0.95,\, 1.5$.
The critical time is $t_c = 0.95$ (black blue curve). The values of
the other parameters are: $x_0=0.1$, $r=2$, $D_1=1.7$.}
\label{fig-8}
\end{figure}

\section{Nonlinear Relaxation Time of the Mean Population Density}

It must be emphasized that to find the time evolution of the mean
population density one can use two different approaches. The first
one was proposed in Ref.~\cite{Jac89}. According to the exact
solution~(\ref{F-7}) of the Verhulst equation~(\ref{F-4}), we can
rewrite this expression in the following form
\begin{equation}
x\left(t\right) = f\left(e^{-rt-L\left(t\right)}\right),
\label{F-17}
\end{equation}
where
\begin{equation}
f\left(q\right)=\left(1+\frac{1-x_{0}}{x_{0}}q(t) \right)^{-1}.
\label{F-18}
\end{equation}
Then, by expanding the smooth function~(\ref{F-18}) in a standard
Taylor power series in $q$ around the point $q=0$ we have
\begin{equation}
f\left(q\right) = \sum\limits_{n=0}^{\infty}\frac{f^{\left(
n\right)} \left(0\right)}{n!}\,q^{n}. \label{F-19}
\end{equation}
After substitution of Eq.~(\ref{F-19}) in (\ref{F-17}) and averaging
we obtain
\begin{equation}
\left\langle x\left(t\right) \right\rangle =
\sum\limits_{n=0}^{\infty} \frac{f^{\left(n\right)} \left(
0\right)e^{-nrt}}{n!}\left\langle e^{-nL\left(t\right)}\right\rangle
\label{F-20}
\end{equation}
or, in accordance with Eq.~(\ref{F-8}),
\begin{eqnarray}
\left\langle x\left(  t\right)  \right\rangle
&=&\sum\limits_{n=0}^{\infty} \frac{f^{\left(  n\right)  }\left(
0\right) e^{-nrt}}{n!}\label{F-21} \\
&\times&\exp\left\{ t\int_{-\infty}^{+\infty}\frac{e^{-nz}-1+n\sin
z}{z^{2}}\,\rho\left( z\right) dz\right\} .\nonumber
\end{eqnarray}
For white Gaussian noise $\xi\left(t\right)$ with kernel
$\rho\left(z\right) = 2D \delta(z)$ we obtain from Eq.~(\ref{F-21})
the following asymptotic series
\begin{equation}
\left\langle x\left(t\right) \right\rangle
=\sum\limits_{n=0}^{\infty}
\frac{f^{\left(n\right)}\left(0\right)}{n!}\,e^{Dtn^{2}-nrt}.
\label{F-22}
\end{equation}
By considering a finite number of terms in this expansion leads to a
wrong conclusion about the critical slowing down phenomenon in such
a system, as found in Ref.~\cite{Leu88}. The exact result is
obtained, of course, by summing all the terms in Eq.~(\ref{F-22}).
Moreover, for most of the kernels $\rho(z)$ the integral in
Eq.~(\ref{F-21}) diverges. Thus, this approach is inappropriate for
our purposes, and it is better to use the direct averaging in
Eq.~(\ref{F-7}). Therefore, using this second approach we have
\begin{equation}
\left\langle x\left(  t\right)  \right\rangle
=\int\nolimits_{-\infty }^{+\infty}\left(
1+\frac{1-x_{0}}{x_{0}}\,e^{-rt-z}\right)  ^{-1}P_{L}\left(
z,t\right)dz. \label{F-23}
\end{equation}

\begin{figure}[h] \vspace{5mm}
\centering{\resizebox{7cm}{!}{\includegraphics{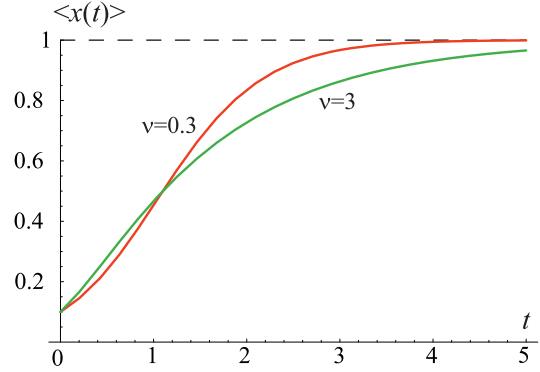}}}
\caption{Nonlinear relaxation of the mean population density in the
case of white shot noise excitation, for three values of the mean
frequency $\nu$, namely $\nu = 0.3, 3$. The values of the other
parameters are: $x_0 = 0.1, r = 2, a_0 = 1$.} \label{fig-9}
\end{figure}

Let us consider now different models of white non-Gaussian noise
$\xi\left(t\right)$. We start with the white shot noise
\begin{equation}
\xi\left(t\right) = \sum\limits_{i}a_{i}\,\delta\left(t-t_{i}\right)
\label{F-24}
\end{equation}
having the symmetric dichotomous distribution of the pulse amplitude
$P(a) = \left[\delta\left(a-a_{0}\right) + \delta
\left(a+a_{0}\right)\right] /2$, mean frequency $\nu$ of pulse
train, and kernel $\rho\left(z\right) = \nu z^{2} P\left(z\right)$.
From Eq.~(\ref{F-8}) we have
\begin{equation}
\left\langle e^{iuL\left(t\right)}\right\rangle = e^{-\nu
t\left(1-\cos a_{0}u\right)}. \label{F-25}
\end{equation}
By making the reverse Fourier transform in Eq.~(\ref{F-25}) we find
the probability distribution of the corresponding L\'{e}vy process
\begin{equation}
P_{L}\left( z,t\right) =e^{-\nu
t}\sum_{n=-\infty}^{+\infty}I_n\left( \nu t\right) \delta \left(
z-na_0\right) , \label{F-26}
\end{equation}
where $I_n\left( x\right) $ is the $n$-order modified Bessel
function of the first kind. The relaxation of the mean population
density $\left\langle x(t)\right\rangle $ is shown in
Fig.~\ref{fig-9}. According to the Eqs.~(\ref{F-23})
and~(\ref{F-26}) the stationary value of the population density in
such a case is $\left\langle x\right\rangle _{st}=1$, but the
relaxation time~(\ref{F-3}) increases with increasing the mean
frequency of pulses.

\begin{figure}[h] \vspace{5mm}
\centering{\resizebox{7cm}{!}{\includegraphics{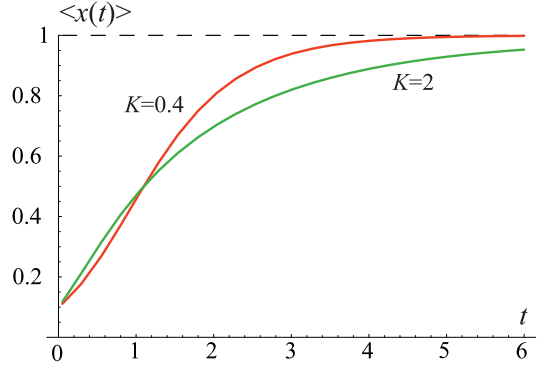}}}
\caption{Nonlinear relaxation of the mean population density in the
case of L\'{e}vy noise with distribution~(\ref{F-13}), for three
values of the parameter $K$, namely $K = 0.4, 2$. The values of the
other parameters are: $x_0 = 0.1, r = 2$.} \label{fig-10}
\end{figure}

For white non-Gaussian noise with the kernel $\rho(z) =
Kz/\left(2\sinh z\right)$ we observe a similar transient dynamics,
which is shown in Fig.~\ref{fig-10}. We have the same stationary
value $\left\langle x\right\rangle_{st}$, and the relaxation time
$T$ increases with increasing the parameter $K$, which is
proportional to the noise intensity.

\begin{figure}[h] \vspace{5mm}
\centering{\resizebox{7cm}{!}{\includegraphics{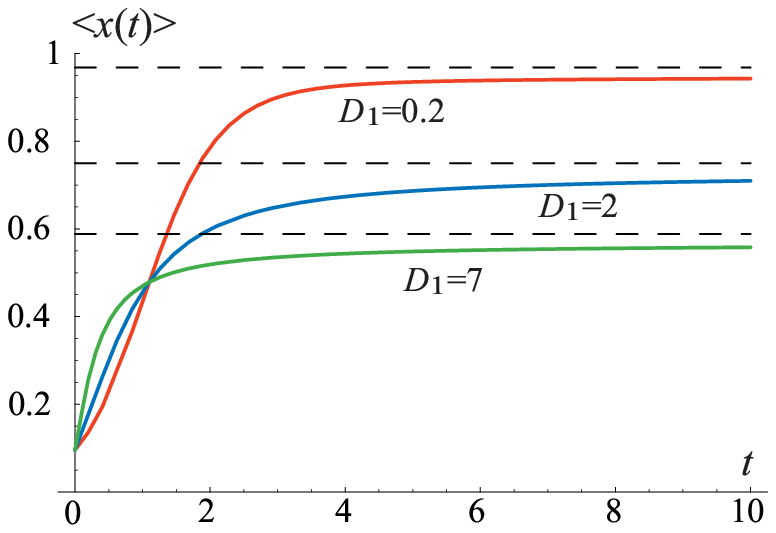}}}
\caption{Nonlinear relaxation of the mean population density in the
case of white Cauchy noise, for three values of the noise intensity
$D_1$, namely $D_1 = 0.2, 2, 7$. The values of the other parameters
are: $x_0 = 0.1, r = 2$.} \label{fig-11}
\end{figure}

Finally, in the case of white Cauchy noise $\xi\left(t\right)$ we
obtain interesting exact analytical results. First of all,
substituting Eq.~(\ref{F-13}) in (\ref{F-23}) and changing the
variable $z=D_{1}ty$ under the integral, we obtain
\begin{equation}
\left\langle x\left(t\right) \right\rangle = \frac{1}{\pi}\int
\nolimits_{-\infty}^{+\infty}\left[
1+\frac{1-x_{0}}{x_{0}}e^{-t\left( r+D_{1}y\right)}\right]
^{-1}\frac{dy}{1+y^{2}}\,. \label{F-27}
\end{equation}
For the stationary mean value $\left\langle x\right\rangle _{st}$ we
find from Eq.~(\ref{F-27})
\begin{equation}
\left\langle x\right\rangle
_{st}=\lim_{t\rightarrow\infty}\left\langle x\left(  t\right)
\right\rangle =\frac{1}{\pi}\int\nolimits_{-\infty
}^{+\infty}\frac{1\left(  r+D_{1}y\right)  dy}{1+y^{2}}\,,
\label{F-28}
\end{equation}
where $1\left(x\right)$ is the step function. After evaluation of
the integral in Eq.~(\ref{F-28}) we obtain finally
\begin{equation}
\left\langle x\right\rangle
_{st}=\frac{1}{2}+\frac{1}{\pi}\arctan\frac {r}{D_{1}}\,.
\label{F-29}
\end{equation}
As it is seen from Fig.~\ref{fig-11} and Eq.~(\ref{F-29}), for small
noise intensity $D_{1}$, with respect to the value of the rate
parameter $r = 2$, the stationary mean value of the population
density is approximately $1$, as for the other white non-Gaussian
noise excitations considered. But for large values of $D_{1}$, this
asymptotic value, which is independent from the initial value of
population density $x_{0}$, tends to $0.5$.

\begin{figure}[h] \vspace{5mm}
\centering{\resizebox{7cm}{!}{\includegraphics{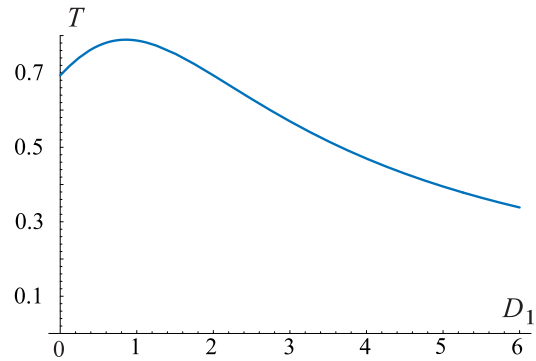}}}
\caption{Nonmonotonic behavior of the nonlinear relaxation time $T$
as a function of the white Cauchy noise intensity $D_1$. The values
of the other parameters are: $x_0 = 0.5, r = 2$.} \label{fig-12}
\end{figure}

It is interesting also to analyze, for this case of white Cauchy
noise, the dependence of the relaxation time $T$ from the noise
intensity $D_{1}$. Substituting Eq.~(\ref{F-27}) in (\ref{F-3}) and
changing the order of integration, for initial condition $x_0 =
0.5$, we are able to calculate analytically the double integral in
$t$ and in $y$ obtaining the final result
\begin{equation}
T=\frac{\pi\ln2}{r\left(  1+D_{1}^{2}/r^{2}\right)
\mathrm{arccot}(D_{1}/r)}\,. \label{F-30}
\end{equation}
We find a nonmonotonic behavior of the relaxation time $T$ versus
the noise intensity $D_{1}$ with a maximum at the noise intensity
$D_1 = 0.75$, as shown in Fig.~\ref{fig-12}.
This nonmonotonic behavior is also visible for another initial
position $x_{0}=0.1$ in Fig.~\ref{fig-11}. Here the relaxation time
to reach the stationary value of mean population density
$\left\langle x\right\rangle_{st}$ increases from very low noise
intensity ($D_1 = 0.2$) to moderate low intensity ($D_1 = 2$), while
decreases for higher noise intensities ($D_1 = 7$). This is also due
to the dependence of $\left\langle x\right\rangle _{st}$ from the
noise intensity $D_1$ (see Eq.~(\ref{F-29})). We note that this
nonmonotonic behavior of the relaxation time $T$ is related to the
peculiarities of the transient dynamics of the mean population
density and it will be object of further investigations.

\section{Conclusions}\label{conc}
In this contribution we have studied the effects of random
fluctuations, i.e. noise, in the vibrational communications
occurring during the mating of \emph{N. viridula}, i.e. the green
bug. In our experimental work we analyzed the behavioural response
of different individuals of \emph{N. viridula} to a deterministic
signal (calling song), measuring for these individuals the the
threshold of the neural activation. Afterwards, we analyzed the
green bug response when a sub-threshold deterministic signal is
added with an external noise source. By using the
\emph{Source-Direction Movement ratio} as indicator of positive
response to the external signal, we observe that the behavioural
activation of the insects is characterized by a non-monotonic
behaviour as a function of the noise intensity $D$, with a maximum
at $D=D_{opt}\approx 1.30 \cdot 10^{-5}$ $V^{2}$. The value
$D_{opt}$ maximizes the efficiency of the sexual communication among
individuals of \emph{Nezara viridula (L.)}, and therefore represents
the optimal noise intensity during the mating behaviour of these
insects. The non-monotonic behaviour observed in the insect response
as a function of the noise intensity is the signature of the
threshold stochastic resonance (TSR)~\cite{Gree04}. By using a
threshold model we obtained numerical results for the threshold
crossing, which corresponds in our model to the behavioural
activation, finding a theoretical value for the optimal noise
intensity. Experimental and numerical results are compared, finding
a good agreement between the values of the optimal noise intensity
obtained by the experimental work and model (see
Figs.~\ref{fig:2}(b) and~\ref{fig:5}).

We analyzed the influence of an external oscillating driving field
on the translocation dynamics of short polymers embedded in a noisy
environment. We simulate the translocation process by letting the
polymer to cross a potential barrier starting from a metastable
state, in the presence of thermal fluctuations. The mean
translocation time as a function of the frequency of the driving
force shows a nonmonotonic behaviour, with the noise intensity
acting as a scaling factor of the values of the crossing times. The
forcing periodic electric field jointly with the temperature of the
system can be able to speed up or slow down the polymer
translocation. In this view, the oscillating electric field
constitutes a tuning mechanism to select a suitable translocation
time of the polymer. This feature may have important biological
effects on the cell metabolism, for example, during a cancer
targeted therapy.

Finally, the transient dynamics of the Verhulst model, perturbed by
arbitrary non-Gaussian white noise, has been investigated. This
well-known equation is an appropriate ecological and biological
model to describe closed-population dynamics, self-replication of
~macromolecules under constraint, cancer growth, spread of viral
epidemics, etc... By using the properties of the infinitely
divisible distribution of the generalized Wiener process, we
analyzed the effect of different non-Gaussian white sources on the
nonlinear relaxation of the mean population density and on the time
evolution of the probability distribution of the population density.
We obtain exact results for the nonstationary probability
distribution in all cases investigated and for the Cauchy stable
noise we derive the exact analytical expression of the nonlinear
relaxation time. Due to the presence of a L\'{e}vy multiplicative
noise, the probability distribution of the population density
exhibits a transition from a trimodal to a bimodal distribution in
asymptotics. This transition, characterized by the appearance of a
minimum, happens at a critical time $t_c$, which can be roughly
evaluated as $t_c \sim 1/D_1$ (where $D_1$ is the noise intensity)
and exactly evaluated from the condition~(\ref{F-15}). Finally a
nonmonotonic behavior of the nonlinear relaxation time of the
population density as a function of the Cauchy noise intensity was
found.

\section*{Acknowledgments}
\label{aknow}
Authors acknowledge the financial support by MIUR.

\label{lastpage-01}


\begin{thebibliography}{999}


\bibitem{Agu01}
N.V. Agudov, B. Spagnolo, \emph{Phys. Rev. E} \textbf{64}, 035102(R)
(2001); N. V. Agudov, A. A. Dubkov, B. Spagnolo, \emph{Physica A}
\textbf{325}, 144 (2003); A. A. Dubkov, N. V. Agudov and B.
Spagnolo, \emph{Phys. Rev. E} \textbf{69}, 061103 (2004); A.
Fiasconaro and B. Spagnolo, \emph{Phys. Rev. E} \textbf{80}, 041110
(6) (2009).

\bibitem{Spa03}
B. Spagnolo, A. Fiasconaro, D. Valenti, \emph{Fluct. Noise Lett.}
\textbf{3}, L177 (2003).

\bibitem{Val04}
D. Valenti, A. Fiasconaro, B. Spagnolo, Physica A 331 (2004)
477-486; B. Spagnolo, M. Cirone, A. La Barbera and F. de Pasquale,
\emph{Journal of Physics: Condensed Matter} \textbf{14}, 2247
(2002).

\bibitem{Zim99}
C. Zimmer, \emph{Science} \textbf{284}, 83 (1999);

\bibitem{Bjo01}
O. N. Bj{\o}rnstad and B. T. Grenfell, \emph{Science} \textbf{293},
638 (2001).

\bibitem{Gre98}
B. T. Grenfell, K. Wilson, B. F. Finkenst\"{a}dt, T. N. Coulson, S.
Murray, S. D. Albon, J. M. Pemberton, T. H. Clutton-Brock, M. J.
Crawley, \emph{Nature} \textbf{394}, 674 (1998).

\bibitem{Chi08}
O. A. Chichigina, \emph{Eur. Phys. J. B} \textbf{65}, 347 (2008).

\bibitem{Giu09}
A. Giuffrida, D. Valenti, G. Ziino, B. Spagnolo, A. Panebianco,
\emph{Eur. Food Res. Technol.} \textbf{228}, 767 (2009); E.
Korobkova, T. Emonet, J.M. Vilar, T.S. Shimizu, P. Cluzel,
\emph{Nature} \textbf{428}, 574 (2004).

\bibitem{Piz09a}
N. Pizzolato, D. Valenti, D. Persano Adorno, B. Spagnolo,
\emph{Cent. Eur. J. Phys.} \textbf{7}, 541 (2009); I. Roeder, M.
Horn, I. Glauche, A. Hochhaus, M.C. Mueller, M. Loeffler,
\textit{Nature Medicine} \textbf{12}, 1181 (2006); Y. Brumer, F.
Michor, E.I. Shakhnovich, \textit{J. Theor. Biol.} \textbf{241}, 216
(2006); F. Michor, Y. Iwasa, and M. Nowak, Proc. Natl. Acad. Sci.
USA \textbf{103}, 14931 (2006); N.L. Komarova, D. Wodarz,
\textit{Theor. Popul. Bio.} \textbf{72}, 523 (2007); N.L. Komarova,
D. Wodarz, \emph{PLos ONE} \textbf{2},e990 (2007); V.P. Zhdanov,
\textit{Eur. Biophys. J.} \textbf{37}, 1329 (2008).

\bibitem{Bra94}
Braun H.A. et al., Nature \textbf{367}, (1994) 270-273.

\bibitem{Mos94}
Moss F., Pierson D. and O'Gorman D., Int. J. of Bifurcation and
Chaos \textbf{4} (6), (1994) 1383-1397.

\bibitem{Gin95}
Gingl Z., Kiss L. B. and Moss F., Europhys. Lett. \textbf{29} (3),
(1995) 191-196.

\bibitem{Pei95}
Pei X., Bachmann K. and Moss F., Phys. Lett. A \textbf{206}, (1995)
61-65.

\bibitem{Pik97}
Pikovsky A.~S., and Kurths J., Phys. Rev. Lett. \textbf{78}, (1997)
775-778.

\bibitem{Noz98}
Nozaki D., Yamamoto Y., Phys. Lett. A \textbf{243}, (1998) 281-287.

\bibitem{Lon98}
Longtin A., Chialvo D.~R., Phys. Rev. Lett. \textbf{81}, (1998)
4012-4015.

\bibitem{Sto00}
Stocks N. G., Phys. Rev E \textbf{64}, (2001) 030902(4); \emph{id.},
Phys. Rev E \textbf{63},(2001) 041114 (9); \emph{id.}, Phys. Lett. A
\textbf{279}, (2001) 308-312; \emph{id.}, Phys. Rev. Lett.
\textbf{84}, (2000) 2310-2314.

\bibitem{Wie94}
Wiesenfeld K. \emph{et al.}, Phys. Rev. Lett. \textbf{72}, (1994)
2125-2129.

\bibitem{Gam95}
Gammaitoni L., Phys. Rev. E \textbf{52}, (1995) 4691-4698; Phys.
Lett. A \textbf{74}, (1995) 315-322.

\bibitem{Wan00}
Wannamaker R. A., Lipshitz S. P., and Vanderkooy J., Phys. Rev. E
\textbf{61}, (2000) 233-236.

\bibitem{Lin04}
Lindner B., Ojalvo J. G., Neiman A., Schimansky-Geier L., Physics
Reports \textbf{392}, (2004) 321-424.

\bibitem{Dua08}
Duarte J. R. R., Vermelho M. V. D., Lyra M. L., Physica A
\textbf{387} (2008) 14461454.

\bibitem{Pol05}
Pankratova E. V., Polovinkin A. V., Spagnolo B., Physics Letters A
\textbf{344} (2005) 43-50; B. Spagnolo \emph{et al.}, Acta Physica
Polonica B, \textbf{38} (5), (2007) 1925-1950.

\bibitem{Spa04}
Spagnolo B., Valenti D., Fiasconaro A., Math Biosci. Eng. \textbf{1}
(2004) 185-211.

\bibitem{Benzi}
Benzi R., Sutera A., Vulpiani A., J. Phys.: Math Gen. \textbf{14}
(1981) L453-L457; Benzi R., Parisi G., Sutera A., Vulpiani A.,
Tellus \textbf{34} (1982) 10-16.

\bibitem{Gam98}
Gammaitoni L., H$\ddot{a}$nggi P., Jung P., Marchesoni F., Rev. Mod.
Phys. \textbf{70}, (1998) 223-287.

\bibitem{Man94}
Mantegna R. N., Spagnolo B., Phys. Rev. E \textbf{49}, (1994)
R1792-R1795; R. N. Mantegna, B. Spagnolo and M. Trapanese, Phys.
Rev. E \textbf{63}, (2001) 011101 (8).

\bibitem{Agu10}
N. V. Agudov, A. V. Krichigin, D. Valenti, and B. Spagnolo, Phys.
Rev. E \textbf{81}, 051123 (8) (2010).

\bibitem{Vil98}
Vilar J. M., Gomila G. and Rubi J. M., Phys. Rev. Lett. \textbf{81}
(1998) 14-17.

\bibitem{Bul91}
Longtin A., Bulsara A., Moss F., Phys. Rev. Lett., \textbf{67},
(1991) 656-659; Bulsara A., Jacobs E. W., Zhou T., Moss F., Kiss L.,
J. Theor. Biol., \textbf{152}, (1991) 531-555; Chialvo D. R.,
Apkarian A. V., J. Stat. Phys., \textbf{70}, (1993) 375-391.

\bibitem{Nei02}
Neiman A., Russell D., Phys. Rev. Lett. \textbf{88}, (2002)
138103(4).

\bibitem{Bah02}
Bahar S., Neiman A., Wilkens L., Moss F., Phys. Rev. E, \textbf{65},
(2002) 050901(R).

\bibitem{Dou93}
Douglass J.K. et al., Nature, \textbf{365} (1993) 337-340.

\bibitem{Rus99}
Russell D. F., Wilkens L. A., Moss F., Nature 402, (1999) 291-294.

\bibitem{Fre02}
Freund J., Schimansky-Geier L., Beisner B., Neiman A., Russell D.,
Yakusheva T. and Moss F., Journal of Theoretical Biology,
\textbf{214}, (2002) 71-83.

\bibitem{Gre00}
Greenwood P. E., Ward L. M., Russell D. F., Neiman A. and Moss F.,
Phys. Rev. Lett. \textbf{84}, (2000) 4773-4776.

\bibitem{Gai97}
P. C. Gailey, A. Neiman, J. J. Collins and F. Moss, Phys. Rev. Lett.
\textbf{79}, 4701 (1997)

\bibitem{Cok99}
$\check{C}$okl A., Virant Doberlet M., McDowell A., Anim. Behav.
\textbf{58}, (1999) 1277-1283.

\bibitem{Cok03}
$\check{C}$okl A., Virant Doberlet M., Annual Review of Entomology
\textbf{48}, (2003) 29-50.

\bibitem{Cok07}
$\check{C}$okl A., Zorovi$\acute{c}$ M., Millar J. G., Behavioural
Processes \textbf{75}, (2007) 4054.

\bibitem{Tod89}
Todd J. W., Annu. Rev. Entomol. \textbf{34}, (1989) 273-292.

\bibitem{Pan00}
Pannizzi A. R., Anais Soc. Entomol. Brasil \textbf{29}, (2000) 1-12.

\bibitem{Bor87}
Borges M., Jepson P. C., Howse P.E., Entomologia Experimentalis et
Applicata \textbf{44}, (1987) 205-212.

\bibitem{Kir64}
Kiritani K., Jpn. J. Appl. Entomol. Zool. \textbf{8}, (1964) 45-53.

\bibitem{Tre81}
Tremblay E., \textit{Entomologia applicata, Vol. 2, parte prima},
(Liguori, 1981) 66.

\bibitem{Fuc03}
Fucarino, \textit{Semiochemical relationships in the tritrophic
system Leguminous, Nezara viridula (L.) and Trissolcus basalis
(Woll.)} (Ph.D. thesis, University of Palermo, Italy, 2003).

\bibitem{Bag08}
Bagwell G. J., $\check{C}$okl A., Millar J. G., Ann. Entomol. Soc.
Am. \textbf{101(1)}: (2008) 235-246.

\bibitem{Cok00}
$\breve{C}$okl A., Virant Doberlet M., Stritih N., Physiol. Entomol.
\textbf{25}, (2000) 196-205.

\bibitem{Higgins2007}
C. F. Higgins, {\it Nature} {\bf 446}, 749 (2007).

\bibitem{Hal2009}
S. Halwachs, I. Sch$\rm{\ddot{a}}$fer, P. Seibel, and W. Honscha,
{\it Leukemia} {\bf 23}, 1087 (2009).

\bibitem{Mann06}
J. T. Mannion, C. H. Reccius, J. D. Cross, and H. G. Craighead, {\it
Biophys. J.} {\bf 90}, 4538 (2006).

\bibitem{Sunda2008}
V. B. Sundaresan and D. J. Leo, {\it Sens. Actuators B: Chem.} {\bf
131}, 384 (2008).

\bibitem{Kas96}
J. J. Kasianowicz, E. Brandin, D. Branton, and D. W. Deamer, {\it
Proc. Natl. Acad. Sci. USA} {\bf 93}, 13770 (1996).

\bibitem{Ake99}
M. Akeson, D. Branton, J. J. Kasianowicz, E. Brandin, and D. W.
Deamer, {\it Biophys. J.} {\bf 77}, 3227 (1999).

\bibitem{Mel00}
A. Meller, L. Nivon, E. Brandin, J. A. Golovchenko, and D. Branton,
{\it Proc. Natl. Acad. Sci. USA} {\bf 97}, 1079 (2000).

\bibitem{Mel02}
A. Meller and D. Branton, {\it Electrophoresis} {\bf 23}, 2583
(2002).

\bibitem{Den03}
J. Deng, K. H. Schoenbach, E. S. Buescher, P. S. Hair, P. M. Fox, S.
J. Beebe, {\it Biophysical J.} {\bf 84}, 2709 (2003).

\bibitem{Ver04}
P. T. Vernier, Y. Sun, L. Marcu, C. M. Craft, M. A. Gundersen, {\it
Biophysical J.} {\bf 86}, 4040 (2004).

\bibitem{Sig08}
G. Sigalov, J. Comer, G. Timp, and A. Aksimentiev, {\it Nano
Letters} {\bf 8}, 56 (2008).

\bibitem{Lat09}
D. K. Lathrop, G. A. Barrall, E. N. Ervin, M. G. Keehan, M. A.
Krupka, R. Kawano, H. S. White, A. H. Hibbs, {\it Biophysical J.}
{\bf 96}, 647a (2009).

\bibitem{Nik09}
A. Nikolaev, and M. Gracheva, {\it Biophysical J.} {\bf 96}, 649a
(2009).

\bibitem{Lub99}
D. K. Lubensky, and D. R. Nelson, {\it Biophys. J.} {\bf 77}, 1824
(1999).

\bibitem{Sto05}
A. J. Storm , J. Chen, H. Zandbergen, and C. Dekker, {\it Phys. Rev.
E} {\bf 71}, 51903 (2005).

\bibitem{For07}
C. Forrey and M. Muthukumar, {\it J. Chem. Phys.} {\bf 127}, 015102
(2007).

\bibitem{Luo08}
K. F. Luo, T. Ala-Nissila, S. C. Ying, and A. Bhattacharya, {\it
Phys. Rev. Lett.} {\bf 100}, 58101 (2008).

\bibitem{Gra08}
M. E. Gracheva, and J. P. Leburton, {\it J. Comput. Electron.} {\bf
7}, 6 (2008).

\bibitem{Piz08}
N. Pizzolato, A. Fiasconaro, B. Spagnolo, {\it Int. J. Bifurc.
Chaos} {\bf 18}, 2871 (2008).

\bibitem{Pan08}
D. Panja, and G. T. Barkema, {\it Biophysical J.} {\bf 94}, 1630
(2008).

\bibitem{Piz09}
N. Pizzolato, A. Fiasconaro, and B. Spagnolo, {\it J. Stat. Mech:
Theory and Exp.} P01011 (2009).

\bibitem{Doe92}
C.R. Doering, and J. C. Gadoua, {\it Phys. Rev. Lett.} {\bf 69},
2318 (1992).

\bibitem{Bie93}
M. Bier, and R.D. Astumian, {\it Phys. Rev. Lett.} {\bf 71}, 1649
(1993).

\bibitem{Bog98}
M. Boguna, J. M. Porra, J. Masoliver, and K. Lindenberg, {\it Phys.
Rev. E} {\bf 57}, 3990 (1998).

\bibitem{Man00}
R. N. Mantegna and B. Spagnolo, {\it Phys. Rev. Lett.} {\bf 84},
3025 (2000).

\bibitem{Dub04}
A. A. Dubkov, N. V. Agudov, and B. Spagnolo, {\it Phys. Rev. E} {\bf
69}, 061103 (2004).

\bibitem{Spa07}
B. Spagnolo, A. A. Dubkov, A. L. Pankratov, E. V. Pankratova, A.
Fiasconaro, A. Ochab-Marcinek, {\it Acta Phys. Pol. B} {\bf 38},
1925 (2007).

\bibitem{Park98}
P. J. Park, and W. Sung, {\it Int. J. Bifurc. Chaos} {\bf 8}, 927
(1998).

\bibitem{Rou53}
P. E. J. Rouse, {\it J. Chem. Phys.} {\bf 21}, 1272 (1953).

\bibitem{Gree04}
Greenwood P. E., M$\ddot{u}$ller U. U., Ward L. M., Phys. Rev. E
\textbf{70}, (2004) 051110 1-10.

\bibitem{Col04} Colazza S., Fucarino A., Peri E., Salerno G., Conti E.,
Bin F., J. Exp. Biol. \textbf{207}, (2004) 47-53.

\bibitem{Cok05}
$\check{C}$okl A., Zorovi$\acute{c}$ M., $\check{Z}$uni$\check{c}$
A., Virant Doberlet M., J. Exp. Biol. \textbf{208}, (2005)
1481-1488.

\bibitem{Pizzolato2010} 
N. Pizzolato, A. Fiasconaro, D. Persano Adorno, and B. Spagnolo, “Resonant activation
in polymer translocation: new insights into escape dynamics of molecules driven by an
oscillating field”, Physical Biology, 7 (2010) 034001-5, doi:10.1088/1478-3975/7/3/034001.

\bibitem{Hor84}
W. Horsthemke and R. Lefever, \emph{Noise-Induced Transitions:
Theory and Applications in Physics, Chemistry and Biology},
(Springer--Verlag, Berlin, 1984).

\bibitem{Eig79}
M. Eigen and P. Schuster, \emph{The Hypercycle: A Principle of
Natural Self-Organization}, (Springer, Berlin, 1979).

\bibitem{Mor82}
A. Morita, J. Chem. Phys. \textbf{76}, (1982) 4191--4194.

\bibitem{Ciu93}
S. Ciuchi, F. de Pasquale, and B. Spagnolo, Phys. Rev. E
\textbf{47}, (1993) 3915--3926.

\bibitem{Mat00}
J. H. Mathis and T. R. Kiffe, \emph{Stochastic Population Models: A
Compartmental Perspective}, (Springer--Verlag, Berlin, 1984).

\bibitem {Eig71}
M. Eigen, Naturwissenschaften \textbf{58}, (1971) 465--523.

\bibitem{Ace06}
L. Acedo, Physica A \textbf{370}, (2006) 613--624.

\bibitem{Bao03}
Bao-Quan Ai, Xian-Ju Wang, Guo-Tao Liu, and Liang-Gang Liu, Phys.
Rev. E \textbf{67}, (2003) 022903-1--022903-3.

\bibitem{Der90}
G. DeRise and J. A. Adam, J. Phys. A: Math. Gen. \textbf{23}, (1990)
L727S--L731S.

\bibitem{Ciu96}
S. Ciuchi, F. de Pasquale, and B. Spagnolo, Phys. Rev. E
\textbf{54}, (1996) 706--716.

\bibitem {McN74}
K. J. McNeil and D. F. Walls, J. Stat. Phys. \textbf{10}, (1974)
439--448.

\bibitem{Oga83}
 H. Ogata, Phys. Rev. A \textbf{28}, (1983) 2296--2299.

\bibitem{Sch72}
F. Schl$\ddot{o}$gl, Z. Phys. \textbf{253}, (1972) 147--161.

\bibitem{Cha76}
S. Chaturvedi, C. W. Gardiner, and D. F. Walls, Phys. Lett. A
\textbf{57}, (1976) 404--406.

\bibitem{Gar77}
C. W. Gardiner and S. Chaturvedi, J. Stat. Phys. \textbf{17}, (1977)
429--468.

\bibitem{Bou82}
V. Bouch$\acute{e}$, J. Phys. A: Math. Gen. \textbf{15}, (1982)
1841--1848.

\bibitem{Leu87}
H. K. Leung, J. Chem. Phys. \textbf{86}, (1987) 6847--6851.

\bibitem{Das83}
A. K. Das, Can. J. Phys. \textbf{61}, (1983) 1046--1049.

\bibitem{Her72}
R. Herman and E. W. Montroll, Proc. Natl. Acad. Sci. U.S.A.
\textbf{69}, (1972) 3019--3023.

\bibitem{Mon78}
E. W. Montroll, Proc. Natl. Acad. Sci. U.S.A. \textbf{75}, (1978)
4633--4637.

\bibitem{Leu88}
H. K. Leung, Phys. Rev. A \textbf{37}, (1988) 1341--1344.

\bibitem{Jac89}
P. J. Jackson, C. J. Lambert, R. Mannella, P. Martano, P. V. E.
McClintock, and N. G. Stocks, Phys. Rev. A \textbf{40}, (1989)
2875--2878.

\bibitem{Bin73}
K. Binder, Phys. Rev. B \textbf{8}, (1973) 3423--3436.

\bibitem{Gol03}
J. Golec and S. Sathananthan, Math. Comput. Modell. \textbf{38},
(2003) 585--593.

\bibitem{Man90}
R. Mannella, C. J. Lambert, N. G. Stocks, and P. V. E. McClintock,
Phys. Rev. A \textbf{41}, (1990) 3016--3020.

\bibitem{Cal07}
H. Calisto and M. Bologna, Phys. Rev. E \textbf{75}, (2007)
050103-1--050103-4(R).

\bibitem{Suz82}
M. Suzuki, K. Kaneko, and S. Takesue, Prog. Theor. Phys.
\textbf{67}, (1982) 1756--1775.

\bibitem{Suz82a}
M. Suzuki, S. Takesue, and F. Sasagawa, Prog. Theor. Phys.
\textbf{68}, (1982) 98--115.

\bibitem{Bre82}
L. Brenig and N. Banai, Physica D \textbf{5}, (1982) 208--226.

\bibitem{Mak85}
J. Makino and A. Morita, Progr. Theor. Phys. \textbf{73}, (1985)
1268--1267.

\bibitem{Mor86}
A. Morita and J. Makino, Phys. Rev. A \textbf{34}, (1986)
1595--1598.

\bibitem{Dub05}
A. A. Dubkov and B. Spagnolo, Fluct. Noise Lett. \textbf{5}, (2005)
L267--L274.

\bibitem{Dub08}
Alexander A. Dubkov, Bernardo Spagnolo, and Vladimir V. Uchaikin,
"L\'{e}vy flights Superdiffusion: An Introduction", Intern. Journ.
of Bifurcation and Chaos (2008), in press.

\bibitem{Fel71}
W. Feller, \emph{An Introduction to Probability Theory and its
Applications, Vol. 2} (John Wiley \& Sons, Inc., New York 1971).

\end{thebibliography}
\end{document}